\documentclass{ar2e}
 \usepackage{amssymb,euscript,amsmath,amsfonts}

\begin{document}

\input epsf.tex    
 
\newcommand{\nn}{\nonumber}
\newcommand{\ie}{{\it i.e.}}
\newcommand{\eg}{{\it e.g.}}
\newcommand{\be}{\begin{equation}}
\newcommand{\ee}{\end{equation}}
\newcommand{\bea}{\begin{eqnarray}}
\newcommand{\eea}{\end{eqnarray}}
\def\bZ            {{\bar Z}}

\jname{ Annual Review of Nuclear and Particle Science}
\jyear{2011}
\jvol{61}
\ARinfo{}

\title{ M-Theory and Maximally Supersymmetric Gauge Theories}

\markboth{Neil Lambert}{M-Theory and Supersymmetric Gauge Theories}

\author{Neil Lambert
\affiliation{Theory Division, CERN, 1211-Geneva 23, Switzerland \\ and \\ Department of Mathematics, King's College London, WC2R 2LS, UK
\flushleft
CERN-PH-TH-2012-93
}}


\begin{abstract}
In this informal review for non-specalists we discuss the construction of maximally supersymmetric gauge theories that arise on the worldvolumes branes in String Theory  and M-Theory. Particular focus is made on the relatively recent construction of M2-brane worldvolume theories. In a formal sense,  the existence of  these quantum field theories can be viewed as predictions of  M-Theory. Their construction is therefore a reinforcement of the ideas underlying String Theory and M-Theory. We also briefly discuss the six-dimensional conformal field theory that is expected to arise on M5-branes. The  construction of this theory is 
not only an important open problem for M-Theory but also a significant challenge to our current understanding of quantum field theory more generally.

\end{abstract}

\maketitle

\section{Introduction}

We probably all want to know if there is a mathematical structure that underlies and explains all of the physical world, which has, for hundreds of years, been so accurately described by one and then another set of mathematical laws. The current set, at the fundamental level, consists of two distinct parts: Relativistic Quantum Field Theory and General Relativity. Despite the fact that they both contain the word `Relativity' they  are in fact  worlds apart.
In particular General Relativity is classical, and has not been successfully quantized (yet) whereas the known consistent quantum field theories do not include gravity.

What would be the wish list for such a set of fundamental laws? The following criteria come to mind:
\begin{itemize}
\item[1.] Mathematically consistent
\item[2.] Include gravity coupled to gauge forces
\item[3.] Quantum
\item[4.] Unique
\item[5.] Predictive
\item[6.] Experimentally correct
\end{itemize}
Well it turns out that there is a theory that, as far as we know,  satisfies the first 4 criteria. It struggles with point 5 and point 6 is, to date (and for the foreseeable future), undecided. Thats not bad, in fact it is very good. For a start this wish list is just that, a wish list. In fact the only criteria that seem necessary are 1,2,3 and 6. No one ever promised that the set of laws that govern our Universe are unique (in the sense that there is no other mathematically consistent possibility). Furthermore, if the complete theory of everything was taught to you by some god in a master class, it is not guaranteed that you could then unambiguously determine our Universe within it and not some other set of laws which are so similar that you can't experimentally choose between them. Just as if that same god gave you a perfect map of the universe, including every star, planet and astroid and then gave the homework problem: find home. Of course point 6 is important. It must be the case that somewhere within this complete theory lies the laws that we observe, but again it would be very hard to find them, just as it could be hard to find our solar system if one starts from the Universe as a whole.

So what is this theory? ``String Theory" comes to mind but there are five versions of that and we no longer believe that strings alone form a complete quantum theory. Rather, we now think that the various string theories
are actually five different perturbative expansions of a single theory known as M-Theory. So this is our current best guess for a fundamental theory of everything. Our understanding is not complete but it is compelling and robust. This is quite an achievement and we should take it seriously.

As we have already said, M-Theory is still quite far from from experimental predictions. It is a supersymmetric, eleven-dimensional theory (so it has the maximal amount of  supersymmetry),  but of course it can be compactified to four dimensions in such a way that minimal or even no supersymmetry is preserved. However it appears to admit a huge (some say $10^{500}$, \eg\ see \cite{Douglas:2006es,Denef:2007pq}) number of possible compactifications. This is what makes criteria 5 and 6 so tricky. Indeed String Theory and M-Theory have so many vacua, which are believed to be consistent, that they run the risk of never going away. For example  it might be  possible to work exclusively with these theories, as they can be arbitrarily close to any other consistent theory that someone might come up with, rather like choosing to only work with rational numbers.\footnote{You are asked not to take this point too seriously.} In this sense String Theory and M-Theory are perhaps better thought of as a framework within which to look for consistent theories of our Universe.

In this review we wish to discuss another part of String Theory and M-Theory. Namely the quantum field theories that are associated to the dynamics of $p$-branes, which are extended objects with $p$-spatial dimensions (\eg\ a string is a 1-brane). In particular String Theory contains D$p$-branes for any $p=0,1,2,...,9$ and M-Theory contains M2-branes and M5-branes. According to String Theory and M-Theory there are local quantum field theories that `live' on these branes and govern their low energy dynamics, in the limit that gravity is decoupled. Furthermore these theories must satisfy certain physical requirements due to the geometrical interpretation of branes in a higher dimensional spacetime.

Therefore, in a sense, String Theory and M-Theory, lead to predictions about quantum field theory. For example, long ago, while it was still in its infancy, String Theory predicted the existence of maximally supersymmetric quantum field theories and these were swiftly constructed  \cite{Gliozzi:1976qd,Brink:1976bc}. We will see that M-Theory `predicts' certain strongly coupled, maximally supersymmetric, conformal field theories in three and six dimensions. We should mention that the existence of such theories first arose in \cite{Nahm:1977tg} which  classified all possible superalgebras. However the predictions of String Theory and M-Theory go further in that they imply that there really is an interacting field theory with these superalgbras and also determine various physical properties (such as the vacuum moduli space).  The three-dimensional theories have now been constructed whereas the construction of the 6 dimensional ones remain a current and important open problem.  In this review we will revisit the  construction of maximally supersymmetric Yang-Mills gauge theories and also the relatively recent construction of the three-dimensional conformal field theories. 

Although not experimental predictions, these are predictions that can be tested - much like the many mathematical 'predictions' of String Theory and M-Theory, several of which have subsequently been proven. Verifying them will not tell us that M-Theory is the correct theory of Nature. But it is a successful test of the ideas of M-Theory as a logically complete quantum theory. Furthermore we will see that that there are interesting mathematical objects that play a key role. This review is meant to introduce the field theories that describe the dynamics of multiple branes in String Theory and M-Theory. For a more technical and detailed review of M2-branes and their Chern-Simons-Matter theories the reader should consult \cite{BLMP}. In addition for a discussion of the resulting $AdS_4/CFT_3$ duality from these models see \cite{Klebanov:2009sg} and also \cite{Beisert:2010jr} for a review of integrability of these models.

The rest of this review is organized as follows. In section 2 we will review String Theory and D-branes. We will then also review how branes and U-duality leads to the notion of a single non-perturbative theory known as M-Theory, which contains M2-branes and M5-branes. In section 3 we use symmetry arguments to construct maximally supersymmetric Yang-Mills theories on the worldvolumes of D$p$-branes in String Theory. In section 4 we then use similar ideas to tackle multiple M2-branes in M-Theory. This leads to new maximally supersymmetric field theories in three dimensions with the structure of a Chern-Simons-Matter gauge theory, and commonly known as BLG theories. These are the first theories to be found in the class that we require, however they are limited to only a handful of examples describing two M2-branes. In section 5 we discuss how this construction should be generalized to lagrangians with slightly less supersymmetry and construct the ABJM/ABJ models, which furnish all the desired quantum field theories for an arbitrary number of M2-branes.  Finally in section 6 we will briefly discuss remaining problem of obtaining interacting six-dimensional conformal field theories for multiple M5-branes as well as some closing comments. We would like to note that our presentation, particularly of sections \ref{Dbranessym} and \ref{abjmetc}, is not primarily presented in a  historical way, but rather is aimed at being pedagogical.

\section{From Strings to Branes}

String Theory started out it's life as an attempt to describe the strong interaction between hadrons by simply imagining that they were tied together by a piece of string. This was, and still is, quite respectable as an effective theory. The rise of QCD meant that it was not to be as a fundamental description of hadrons. However by simply adjusting the tension of the string by a mere $40$ orders of magnitude it became clear that String Theory was a fundamental theory of gravity  and gauge forces \cite{Scherk:1974mc}. In particular strings come in two types: open (with end points) and closed (no end points). These strings are then allowed to split and join with an amplitude that is governed by the so-called string coupling constant $g_s$. One can then set up analogues of perturbative Feynman rules that allow one to compute physical processes as an expansion in the number of times the strings split and joint and hence as a power series in $g_s$. It turns out that there are five ways to do this, \ie\ there are five possible String Theories: Type IIA, Type IIB, Type I, heterotic $E_8\times E_8$ and heterotic $SO(32)$. The reader is referred to \cite{Green:1987sp,Green:1987mn,Polchinski:1998rq,Polchinski:1998rr,Becker:2007zj} for an introduction to the many facets of String Theory.

Quantizing a string leads to an infinite tower of spacetime states, corresponding to the possible vibrational modes. Furthermore quantum consistency requires that these   live in ten dimensions. If the string tension is set near the Planck scale then only the lightest, massless states are relevant for particle physics. The infinite tower of states is crucial for the finiteness and UV completeness of String Theory but it doesn't seem so important for phenomenology. 

In particular if we look at the spectrum of closed strings we find gravitons at the massless level (along with some other fields, including a scalar field called the dilaton whose vev determines $g_s$). Therefore quantizing closed strings leads to theories where the metric is dynamical and indeed Einstein's equations come out from the theory. Examining the low energy effective actions of the  massless closed string modes leads to the ten-dimensional supergravity theories \cite{Chapline:1982ww,Giani:1984wc,Huq:1983im,Campbell:1984zc,Schwarz:1983wa,Howe:1983sra}.  Thus gravity is a consequence of the quantum string and is not at all at odds with quantum theory as seems to be the case in standard approaches to quantum theory.

\subsection{D-Branes}

One might think that this was good enough and never discuss open strings (only the Type I string seems to need them). However examining the effective supergravities that describe the low energy dynamics of the closed string modes, one finds interesting states: $p$-branes. These are black hole-like solutions (in particular they are like extremal Reisner-Nordstrom black holes) where the singularity is not a point in space but is extended over a $p$-dimensional surfaces. They are typically non-perturbative soliton-like states whose tension scales as an inverse power of $g_s$. But they are supersymmetric and as such one can trust several calculations that use only the two-derivative supergravity approximation to the full String Theory.  Taking these states seriously is very important. In  doing so one realizes that the five known perturbative String Theories are actually all related and in many cases dual to each other \cite{Hull:1994ys,Witten:1995ex}. In particular what is considered a fundamental string in one theory appears as a soliton $p$-brane state in the dual version. Thus there is a sort of democracy amongst the various branes \cite{Townsend:1995gp} and the fundamental string is only preferred by perturbation theory \cite{Hull:1995xh}.

It turns out that many (but not all) of these $p$-brane states can be described within the tools of perturbative String Theory. In particular in \cite{Polchinski:1995mt} D$p$-branes were identified with a large class of these supergravity $p$-brane solitons (the ones that carry so-called Ramond-Ramond charges). A D$p$-brane is simply a $p$-dimensional surface in space which is the allowed end-point of open strings. In particular the type IIA string has supersymmetric $p$-branes for even $p$ whereas the type IIB string has supersymmetric $p$-branes for odd $p$. The heterotic strings cannot have D$p$-branes and the Type I string has $p=1,5$. In fact, in this language, Type I String Theory is essentially just Type IIB String Theory in the presence of space-filling D9-branes along with a certain projection that is required for consistency and which removes certain states (such as the D3- and D7-branes). 

This definition of D$p$-branes is important as essentially all their  dynamics can be determined by open String Theory, including all the higher modes and couplings to closed strings and hence gravity. It had been known since the earliest days of String Theory that the lightest modes of open strings are vector gauge bosons and quantization of open strings leads to low energy dynamics of non-abelian gauge theories \cite{Paton:1969je}. 

Thus the modern picture of String Theory was born. In this vsiew spacetime is ten-dimensional and filled with closed strings whose quantum fluctuations, splittings and joinings give a UV finite quantum theory of gravity. Inside such a ten-dimensional spacetime one can have all sorts of configurations of D$p$-branes, intersecting in complicated ways, with Yang-Mills gauge and matter fields propagating along their worldvolumes, whose dynamics arises from the quantum splitting and joinings of open strings. Furthermore the open strings can join up to form closed strings, corresponding to the fact that D$p$-branes are massive objects and hence source gravity. Here we can see why there are so many string vacua: essentially any stable configuration of D$p$-branes leads to a low-energy theory with Yang-Mills gauge theories coupled to matter and gravity - many of which could be a suitable home for some observer. The Standard Model does not seem special or pre-designed even though it is clearly just right for us, in much the same way that our solar system is not unique within the Universe, especially constructed in just the right way.

\subsection{M-Theory}

Once one realizes that the various String Theories are related to each other it begs the question of what is the bigger, umbrella, theory that contains them all. This is known as M-Theory. To use a time-honoured analogy consider the story of five blind scientists who are studying an elephant. One feels the left  leg, one the right   leg,  one the ear, another the trunk  and yet another the tail. They all report to each other and describe seemly very different things (although the two working on the legs have established a certain duality). Well in this analogy M-Theory is the elephant and the   scientists are the String Theorists. For a more precise description of M-Theory see \cite{Townsend:1996xj}  and for a review of its branes see \cite{Berman:2007bv} (see also \cite{Obers:1998fb,Acharya:2004qe} for reviews on other aspects of  M-Theory).

However in our case we are a little be luckier. We claim that the whole of M-theory can be deduced by a (non-perturbative) limit of Type IIA String Theory if one properly accounts for the various branes. In particular it has been known for years that the low energy effective action of Type IIA String Theory, called type IIA supergravity, is obtained by the dimensional reduction of eleven-dimensional supergravity \cite{Cremmer:1978km}. Thus the massless modes of the type IIA string are in agreement with the massless Kaluza-Klein fields. Furthermore the fundamental string can be viewed as arising from a 2-brane object  in eleven dimensions that is wrapped on the Kaluza-Klein circle (there is no stringy object in eleven-dimensions since then it could be obtained as a String Theory). 

This is essentially a more modern refinement of older ideas concerning the role of membranes and eleven-dimensional supergravity \cite{Bergshoeff:1987cm,Duff:1987bx}. The important key idea is the realization that the radius of the extra dimension is interpreted in String Theory as the coupling constant \cite{Witten:1995ex}:
\be
R_{11}= g_sl_s\ ,
\ee
where $l_s$ is the string length - that is the scale set by the tension of a fundamental string. Thus by construction Type IIA String Theory is the expansion of eleven-dimensional M-Theory around a spacetime with a zero-sized circle. Turning this around one sees that M-theory can be defined as the  strong coupling limit, $g_s\to\infty$, of Type IIA String Theory. This also implies that there is no natural parameter with which to construct a perturbation expansion. M-Theory is therefore inherently non-perturbative. Furthermore attempts to quantize the 2-brane as we do for fundamental strings fail  \cite{deWit:1988ct,deWit:1988ig}. So to date there is no satisfactory  microscopic definition or construction of M-Theory, we must piece it together like the scientists and their elephant.
 
How do the non-perturbative states match up \cite{Townsend:1995kk}? The Kaluza-Klein momentum modes give rise to D0-branes. This works nicely since a bound state of $k$ D0-branes has a mass $k/g_sl_s = k/R_{11} $ (they are at threshold) and this agrees with a state with $k$ units of Kaluza-Klein momentum. We have already said that the string comes from a wrapped 2-brane state: the M2-brane. Therefore there must be an unwrapped 2-brane state and indeed there is: the D2-brane. Type IIA String Theory also has D4-branes, NS5-Branes (a example of a solitonic brane in String Theory that is not a D-brane), D6-branes and D8-branes. The D4-brane and NS5-brane can be identified with the wrapped and unwrapped states of a 5-brane object in eleven-dimensions. The D6-brane is the electromagnetic dual of the D0-branes and hence is given by a  Kaluza-Klein monopole \cite{Sorkin:1983ns}. Finally this leaves the D8-brane but it turns out that one cannot take a strong coupling limit of D8-branes and stay within the supergravity approximation \cite{Aharony:2010af}.

Thus M-Theory contains just two types of branes: M2-branes and M5-branes. These can also be found as solitonic black hole-like solutions of eleven-dimensional supergravity \cite{Duff:1990xz,Gueven:1992hh}. Following our interpretation of M-Theory as the strong coupling limit of Type IIA String Theory, M2-branes arise as the strong coupling limit of D2-branes and M5-branes arise as the strong coupling limit of D4-branes. Since the worldvolume theories of D2-branes and D4-branes are given by open strings whose dynamics are governed by maximally supersymmetric Yang-Mills gauge theories we learn that the worldvolume theories of M2-branes are determined by the strong coupling (IR) conformal fixed point of maximally supersymmetric three-dimensional super-Yang-Mills and the worldvolume theory of M5-branes by the strong coupling (UV) conformal fixed point of maximally supersymmetric five-dimensional super-Yang-Mills. Thus M-theory predicts the existence of three and six-dimensional  maximally supersymmetric  conformal field theories \cite{Strominger:1995ac,Witten:1995em}\footnote{These six-dimensional theories can also be derived directly from String Theory \cite{Witten:1995zh}.}. The second case is particularly dramatic as five-dimensional Yang-Mills theories are non-renormalizable. But M-theory claims that there is a good UV fixed point and, furthermore, the fixed point conformal field theory is six-dimensional \cite{Seiberg:1997ax}!

Finally we should mention an important point that applies to both the D-brane and M-brane worldvolume theories. As it stands we have been talking about theories of everything, where the branes are described by low energy fluctuations but which also must source gravity and hence curve spacetime. However in all the cases considered here we can take a so-called decoupling limit where we can essentially turn off gravity. This is just what we also do in the Standard Model where we know that ultimately all the quarks and leptons gravitate but these interactions can be safely neglected. Formally to perform this decoupling limit  one lets the Planck length go to zero, while keeping the length scales and energies associated to the brane fluctuations finite. In this way the worldvolume theories of the D-branes and M-branes decouple from gravity and become flat space, Poincar\'e invariant, local quantum field theories.

\section{D-branes and Yang-Mills Gauge Theories with 16 Supersymmetries}
\label{Dbranessym}

Let us look now at how we could derive the effective theory of D-branes. In fact some 35 years ago String Theory first predicted the existence of maximally supersymmetric field theories as the low energy effective action of the lightest open string modes. This led directly to the construction of maximally supersymmetric Yang-Mills \cite{Gliozzi:1976qd,Brink:1976bc}. Here we will essentially re-derive this result from a modern, D-brane, perspective (these theories were first identified with D-branes in \cite{Witten:1995im}). In addition we will use a   different argument. Rather than using the open strings and their dynamics to deduce the low energy effective action, we will deduce it by looking for field theories with the correct symmetries. We should stress that the point of this section is not so much as to promote the claim that String Theory predicts maximally supersymmetric gauge theories as a great success (even though  it did indeed do just this), all this has been known for many years. Rather we wish to present a unified treatment of D-branes that we can then apply to the M2-branes and, hopefully someday,  M5-branes.

So let us consider a stack of $N$ D$p$-brane of type II String Theory in ten-dimensional flat space. The D-brane is stretched along the $x^0,x^1,...,x^p$ directions and hence breaks the Lorentz group: $SO(1,9)\to SO(1,p)\times SO(9-p)$. Here $SO(1,p)$ becomes the Lorentz symmetry group of the field theory on the brane that describes its low energy fluctuations. The $SO(9-p)$ factor can be thought of as coming from rotations of the ${\mathbb R}^{9-p}$ transverse space to the D-branes. This translates into an $R$-symmetry of the effective theory.

Open strings have half as much supersymmetry as closed strings. The simple reason for this is that he left and right moving modes of closed strings are independent whereas the open string boundary conditions require that they are related. This halves the number of oscillators and supersymmetries. In particular, a D$p$-brane preserves the supersymmetries
\be\label{Dpsusy}
\Gamma_{012...p}\epsilon_L = \epsilon_R\ ,
\ee
here $\epsilon_L$ and $\epsilon_R$ are the two spacetime supersymmetry generators obtained from the left and right moving sectors of the closed string. Type IIA strings have $\Gamma_{11}\epsilon_L= \epsilon_L$, $\Gamma_{11}\epsilon_R=-\epsilon_R$ whereas type IIB strings have $\Gamma_{11}\epsilon_{L/R}=\epsilon_{L/R}$. Thus we see that eq.(\ref{Dpsusy})  only has solutions if $p$ is even for type IIA strings and $p$ odd for type IIB strings.

So let us try to construct such a theory for the worldvolume dynamics of $N$ parallel D$p$-branes.  This theory should have $9-p$ scalar fields $X^I$, $I=1,...,9-p$ that represent the fluctuations in the transverse space. These are the Goldstone bosons for broken translational invariance. There are also fermions which follow from supersymmetry, but also arise as Goldstinos for the broken supersymmetries. For simplicity we will simply consider D$p$-branes in type IIA string theory, so in particular $p$ is even. In this case we can write
\be
\epsilon = \epsilon_L+\epsilon_R\ ,
\ee
so that $\epsilon_{L/R} = \frac{1}{2}(1\pm \Gamma_{11})\epsilon$.
Thus  $\epsilon$ is a real, 32 component spinor of $Spin(1,9)$. The supersymmetry constraint eq.(\ref{Dpsusy}) is now
\be\label{Dpsusy2}
\Gamma_{012...p}\Gamma_{11}^{p/2+1}\epsilon = \epsilon\ ,
\ee
leaving 16 independent components. Thus the Goldstinos should satisfy
\be\label{Fsusy}
\Gamma_{012...p}\Gamma_{11}^{p/2+1}\Psi = -\Psi\ .
\ee

Before we write down the supersymmetry transformations we should note that the fields need not just real valued but  rather, in general, should take values in some vector space (a vector space because we will need to take linear combinations and add various fields in order to construct a theory).
Thus we introduce  an index $a$ on the fields, corresponding to expanding the fields in terms of some basis $T^a$ of the vector space, \eg\ $X^I = X^I_aT^a$. Furthermore we need this vector space to have an inner-product, so that we can extract real numbers from the fields:
\be
\langle X^I,X^J\rangle = h^{ab}X_a^IX^J_b\ .
\ee
Henceforth we will use $h^{ab}$ and its inverse $h_{ab}$ to raise and lower indices (for simplicity one can just assume that $h^{ab}=\delta^{ab}$).

We can now construct a trial supersymmetry relation between the scalars and fermions. For the scalars we guess
\be
\delta X^I_a = i\bar\epsilon \Gamma^I\Psi_a\ .
\ee
This is a typical expression, indeed all supersymmetric theories with scalars have such a transformation. We note that, because of eq.(\ref{Dpsusy2}) and eq.(\ref{Fsusy}), $\delta X^I$ is non-vanishing which, for example, forbids inserting a factor of $\Gamma_{11}$ into the right hand side.

However, as we will see,  what ultimately determines the dynamics comes from the fermion supersymmetry transformation. For a free theory one has
\be
\delta \Psi_a = -\Gamma^\mu\Gamma^I\partial_\mu X^I_a\epsilon\ ,
\ee
where $\mu=0,1,2,...,p$. To construct an interacting theory we need to add more terms to the right hand side. What can we add? It needs to be made from the scalar $X^I$ and needs to be $SO(9-p)$ invariant.  Therefore a the simplest guess is
\be
\delta \Psi_a = -\Gamma^\mu\Gamma^I\partial_\mu X^I_a\epsilon - \frac{1}{2}f^{cd}{}_a X_c^IX_d^J\Gamma_{IJ}\Gamma_{11}\epsilon\ .
\ee
Note the appearance of $\Gamma_{11}$, this arises to ensure that $\delta\Psi_a$ respects the conditions eq.(\ref{Dpsusy2}) and eq.(\ref{Fsusy}).
Here $f^{cd}{}_a$ are some constants, which, by construction, are antisymmetric in $c\leftrightarrow d$. 
We could have considered a linear term in $X^I_a$ however this will not lead to an interacting theory; we need something non-linear. We could also try higher order terms but for now a  quadratic term turns out to be enough. 

What is the consequence of this? To find out we need to close the algebra. This means that we need to evaluate 
\be
[\delta_1,\delta_2] X^I_a =  -(2i\bar\epsilon_2\Gamma^\mu \epsilon_1) \partial_\mu X^I_a - f^{cd}{}_a(2i\bar\epsilon_2\Gamma^{J}\Gamma_{11}\epsilon_1 X^J_d)X^I_c\ .
\ee
The first term is just a translation, what we always expect from closure of a supersymmetric theory. The second term is new. Whatever it is, it must be a symmetry of the theory. It has the form
\be
\delta_\Lambda X^I_a =  \Lambda^c{}_a X^I_c\ ,
\ee
where $\Lambda^c{}_a=-2i\bar\epsilon_2\Gamma^{J}\Gamma_{11}\epsilon_1 X^J_df^{cd}{}_a$. Thus $\Lambda^c{}_a$, through its dependence on $X^J_c$ is spacetime dependent. So we have discovered that we must be talking about a gauge theory. 

If we are talking about a gauge theory then we need a gauge field and hence we introduce a gauge covariant derivative
\be
D_\mu X^I_a = \partial_\mu X^I_a - A_\mu{}^b{}_aX_b^I\ .
\ee
This must satisfy the usual properties of a gauge connection. In particular we can construct the field strength $F_{\mu\nu}{}^b{}_a$ through:
\be
[D_\mu,D_\nu] = F_{\mu\nu}{}^b{}_aX^I_b\ .
\ee\
Next we need to come up with a supersymmetric transformation for $A_\mu{}^b{}_a$. A natural guess is 
\be
\delta A_\mu{}^b{}_a = i\bar\epsilon \Gamma_\mu \Gamma_{11}
\Psi_c f^{cb}{}_a\ ,
\ee
again the appearance of $\Gamma_{11}$ ensures that  eq.(\ref{Dpsusy2}) and eq.(\ref{Fsusy}) are respected.
With all this in place we need to re-think $\delta \Psi_a$, which we said determined everything.  We now see that the most natural guess is
\bea
\delta \Psi_a &=& \frac{1}{2}F_{\mu\nu}{}^c{}_dh_{ab}f^{bd}{}_c\Gamma^{\mu\nu}\Gamma_{11}\epsilon -\Gamma^\mu\Gamma^ID_\mu X^I_a\epsilon - \frac{1}{2}f^{cd}{}_a X_c^IX_d^J\Gamma_{IJ}\Gamma_{11}\epsilon\ \nn\\
\delta A_\mu{}^b{}_a &=& i\bar\epsilon \Gamma_\mu \Gamma_{11}
\Psi_c f^{cb}{}_a\ \\
\delta X^I_a &=& i\bar\epsilon \Gamma^I\Psi_a \nn\ ,
\eea
where for completeness we have listed the supersymmetry transformations for all the fields.

Carrying through all the calculations one sees that this works. Namely the transformations close, on-shell, on to translations and gauge transformations.
However we do encounter one constraint:
\be\label{Jacobi}
f^{ab}{}_df^{cd}{}_e +f^{bc}{}_df^{ad}{}_e+f^{ca}{}_df^{bd}{}_e =0 \ .
\ee
We immediately recognize this as the Jacobi identity for a Lie algebra with generators $T^a$ and Lie bracket
\be
[T^a,T^b] = if^{ab}{}_cT^c\ .
\ee
We also assume that the metric is the standard Killing metric $h^{ab} = -f^{ac}{}_d f^{bd}{}_c$.
The Jacobi identity has a well-known interpretation, namely it ensures that the adjoint map $ad_A(X) = i[A,X] $ acts as a derivation of the Lie-algebra:
\be
ad_A[X,Y] = [ad_A(X),Y] + [X,ad_A(Y)]\ .
\ee
It also ensures that the map $A\to ad_A$ is a representation of the Lie algebra where the Lie algebra itself is the vector space on which the representation acts.

The fact that the supersymmetry algebra only closes on-shell is helpful here, since this means that the equations of motion are determined for us. One then learns that these equations of motion are derived from an action that is invariant under supersymmetry. So for the sake of brevity let us just give the action:
\bea
S_{SYM} &=& -\int d^{p+1}x\Big[ -\frac{1}{4}F_{\mu\nu}{}^a{}_bF_{\mu\nu}{}^b{}_a + \frac{1}{2}D_\mu X^I_a D^\mu X^{aI} +\frac{i}{2}\bar\Psi^a\Gamma^\mu D_\mu \Psi_a \nn\\
&&+\frac{i}{2}\bar\Psi^a\Gamma^I\Gamma_{11} X^I_b\Psi_c f^{bc}{}_a + \frac{1}{4} f^{ab}{}_ef^{cde} X^I_aX^J_bX^I_cX^J_d \Big]\ .
\eea
Note that the action is only invariant if the inner-product is gauge invariant; \eg\  $\delta_\Lambda \langle X^I,X^J\rangle=0$. This in turn implies that
\be
f^{abc}=h^{cd}f^{ab}{}_d=f^{[abc]}\ ,
\ee
which is a well-known property of Lie-algebras.

Where is the coupling constant? Well one sees that the structure constants $f^{ab}{}_c$ can be rescaled and still preserve eq.(\ref{Jacobi}). In the limit that $f^{ab}{}_c\to 0$ we see that the interaction terms all decouple and theory is weakly coupled. From the D$p$-brane perspective one is therefore led to identify 
\be
f^{ab}{}_c = c_p g_s 
l_s^{\frac{3-p}{2}} \hat f^{ab}{}_c\ , 
\ee 
where $\hat f^{ab}{}_c $ are the structure constants in some normalized basis (\eg\ such that ${\rm tr}(T^aT^b) = \delta^{ab}$), $g_s$ is the string coupling constant and $l_s=\sqrt{\alpha'}$ is the string length which is needed on dimensional grounds. We have also allowed for the possibility of a constant of proportionality $c_p$ which can be determined by comparing with string calculations (in which case one finds $c_p= (2\pi)^\frac{p-2}{2}$   \cite{Polchinski:1995mt}).

This is the maximally supersymmetric Yang-Mills gauge theory in $p+1$-dimensions. Although we have only derived it for $p$ even (corresponding to type IIA String Theory) the same result is true when $p$ is odd (in which case $\epsilon$ and $\Psi$ are both eigenstates of $\Gamma_{11}$ with eigenvalue $1$ but $\Gamma_{01..p}\epsilon=\epsilon$ while $\Gamma_{012..p}\Psi=-\Psi$ ). Note that we have not used the minimal spinor representation in $p+1$ dimensions but rather kept the full 32-component spinors of ten dimensions. One can of course decompose $\Psi$ and $\epsilon$ accordingly however we have not done this as we believe that  the current convention shows the relation to spacetime more clearly. In addition, the reduction to minimal $p+1$-dimensional spinors varies depending on the nature of spinor representations for the various values of $p$ which would unduly complicate our analysis.

We have derived the low energy effective action that arises from the quantization of open strings that end on D$p$-branes in the limit that gravity has been decoupled.  It is somewhat remarkable that we have arrived at Yang-Mills theories since we never asked for gauge symmetry. Indeed having started from an essentially gravitational theory,  closed String Theory, we have `deduced' Yang-Mills gauge theories as arising on the worldvolumes of various gravitational solitons; the D$p$-branes.

\subsection{Spacetime Interpretation}

Now that we have constructed theories with the correct symmetries to describe multiple, parallel D$p$-branes in type II string theory, we need to see if their predictions agree with what is expected geometrically.

In particular we need to look at the vacuum moduli space. From the spacetime perspective this corresponds to the set of all configurations that the D$p$-branes can take which preserve all the supersymmetries. As such we expect that the moduli space is that of $N$ indistinguishable objects in ${\mathbb R}^{9-p}$:
\be\label{Vmodspace}
{\cal M} = \frac{({\mathbb R}^{(9-p)})^N}{S_N}\ ,
\ee
where $S_N$ is the symmetric group acting on $N$ objects.

To see that this is indeed the case consider the simplest gauge group: $U(N)$. The Lie algebra is just the set of Hermitian $N\times N$ matrices. Note that   the component of $X^I$ that is proportional to the $N\times N$ identity matrix commutes with all other fields and hence does  not interact with any other field.  This therefore describes the over-all translations of the stack of D$p$-branes in the transverse space.

 The vacuum consists of constant scalars $X^I$ which are mutually commuting:
\be
[X^I,X^J]=0\ .
\ee
 Therefore, up to conjugation, which is just a constant gauge transformation, we can choose all the $X^I$ to be diagonal:
\be\label{Xvac}
X^I  =  \left(\begin{array}{lllll}x^I_1 &&&&\\ &x^I_2&&&\\  &&x_3^I &\\ &&& \ddots &\\ &&&&x_N^I\end{array}\right)\ .
\ee
The set of all $x^I_i$, $i=1,...,N$ is therefore ${\mathbb R}^{(9-p)N}$. We interpret the vector $x^I_i\in {\mathbb R}^{(9-p)}$, for a fixed $i$, to be the position of the ith D$p$-brane in the ${\mathbb R}^{(9-p)}$ transverse space.

However the moduli space of vacua consists of this configuration space, modulo gauge transformations. Thus we need to identify two vacua as equivalent if they differ by a gauge transformation: $X^I \simeq  U X^I U^\dag$, $U\in U(N)$. We have already used such gauge transformations to bring $X^I$ into a diagonal form. However there still remain gauge transformations that act on $X^I$ but   keep it diagonal. To see this we observe that if we chose
\be\label{U2}
U = \left(\begin{array}{lllll}0 &1&&&\\ 1&0&&&\\ &&1&& \\ && & \ddots &\\ &&&&1\end{array}\right)\ ,
\ee
then $U X^I U^\dag$ simply interchanges $x_1^I\leftrightarrow x^I_2$. Clearly one could perform similar transformations that act as $x_i^I\leftrightarrow x^I_j$ for any pair $x^I_i$ and $x^I_j$. Thus subgroup of $U(N)$ that preserves the diagonal form eq.(\ref{Xvac}) is just the symmetric group $S_N$ (which is just the Weyl group of $SU(N)$). Therefore we have indeed shown that the vacuum moduli space is eq.(\ref{Vmodspace}). 

 Of course there are other choices of gauge group. In particular  $SO(2N)$, $SO(2N+1)$ and $Sp(N)$ for any $N$ as well as the five exceptional Lie groups. The three additional infinite families can all be found within String Theory, they arise in the presence of so-called orientifolds. However, so far, the exceptional maximally supersymmetric gauge theories have not been found. Nevertheless, the point we wish to observe is that the maximally supersymmetric gauge theories that String Theory predicts to arise on D$p$-branes have been found (in this case many years ago while String Theory was in it infancy) and have the predicted properties. 

\section{M2-branes and Chern-Simons-Matter Theories: 16 Supersymmetries}

Next we wish to consider the same situation for M-Theory. As we've mentioned there are just two types of branes to consider in M-Theory: M2-branes and M5-branes. Let us consider the M2-branes. The M5-branes remain an important unsolved problem which we will return to in the conclusions.

Unlike the case of D$p$-branes there is no definition of the worldvolume dynamics of M-branes in terms of open strings that end on them. Therefore we cannot deduce their dynamics from first principles as was done in \cite{Gliozzi:1976qd,Brink:1976bc}. However some features of the required M2-brane theories were anticipated in \cite{Schwarz:2004yj,Basu:2004ed}. Here we can follow the discussion of the previous section and try to deduce the effective action by looking for field theories with the correct symmetries. This was done in \cite{Bagger:2006sk,Gustavsson:2007vu,Bagger:2007jr,Bagger:2007vi} and was  successful in that it does lead a class of lagrangian theories with all the right features. This in itself was something as a surprise as it had been thought for 30 years that the only maximally supersymmetric lagrangians were those of super-Yang-Mills that we constructed above. In particular it was previously thought that Chern-Simons theories could have at most 6 supersymmetries \cite{Kao:1992ig}. One of the surprises of these new models is that, although the gauge fields are related to the other fields by supersymmetry, they do not come in the same representation of the gauge group. This is consistent with standard results in supersymmetry because the gauge fields do not carry any independent degrees of freedom. Furthermore the amount of supersymmetry depends on the choice of gauge group, which are typically not simple.

However the story turns out to be more complicated because this class of lagrangians found in  \cite{Bagger:2006sk,Gustavsson:2007vu,Bagger:2007jr,Bagger:2007vi} is too small  and is only capable of describing two M2-branes   (as we will discuss later). Nevertheless it opened the door to a class of highly supersymmetric three-dimensional conformal field theories. Let us review this construction now and discuss the appropriate generalization in the next subsection.

Proceeding as before we note that supersymmetries that are preserved by the M2-branes satisfy
\begin{equation}
\Gamma_{012}\epsilon=\epsilon\;.
\end{equation}
The fermions are then goldstino modes and satisfy $\Gamma_{012}\Psi=-\Psi$ and let us call the scalar fields $X^I$, $I=3,4,..,10$. As with D-branes, we assume that they live in some vector space with a basis $T^a$, {\it e.g.} $X^I = X^I_aT^a$.
We start by noting that the free theory has the supersymmetry transformations
\begin{eqnarray}
\delta X^I_d &=& i\bar\epsilon\Gamma^I\Psi_d \nonumber\\
\delta \Psi_d &=& \partial_\mu X^I_d\Gamma^\mu\Gamma^I\epsilon\ .
\end{eqnarray}
To introduce interactions we need to include a term in $\delta\Psi$ that is non-linear in the scalar fields. Now $\Psi$ and $\epsilon$ have opposite eigenvalues with respect to $\Gamma_{012}$ and in addition $[\Gamma_{012},\Gamma_\mu]=0$ but $\{\Gamma_{012},\Gamma_I\}=0$. Thus any term on the right hand side of $\delta\Psi_s$ must have an odd number of $\Gamma^I$ factors. Furthermore we wish to look for conformal field theory. Since the scaling dimensions of $X^I_a$, $\Psi_a$ and $\epsilon$ are  $\frac{1}{2}$, $1$ and $-\frac{1}{2}$ respectively we see that the interaction term we are looking for should be cubic in $X^I_a$. Thus a natural guess is
\begin{eqnarray}
\delta X^I_d &=& i\bar\epsilon\Gamma^I\Psi_d \nonumber\\
\delta \Psi_d &=& \partial_\mu X^I_a\Gamma^\mu\Gamma^I\epsilon - \frac{1}{3!}X^I_aX^J_bX^K_c f^{abc}{}_d\Gamma^{IJK}\epsilon\ .
\end{eqnarray}
Here we  have introduced coupling constants $f^{abc}{}_d$, which, by construction, are anti-symmetric in the first three indices. One can consider a more general ansatz which has terms that are not totally antisymmetric but, assuming scale invariance, this will not work \cite{Bagger:2008se}. However one can add massive deformations that are consistent with all 16 supersymmetries \cite{Gomis:2008cv,Hosomichi:2008qk}, 
corresponding to turning on background supergravity flux \cite{Lambert:2009qw}.

Next we must check that this superalgebra closes. If we compute $[\delta_1,\delta_2]X^I_a$ we find
\begin{equation}
[\delta_a,\delta_2]X^I_d = -2i\bar\epsilon_2 \Gamma^\mu \epsilon_1 \partial_\mu X^I_d - (2i\bar\epsilon_2\Gamma^{JK}\epsilon_1 X^J_aX^K_bf^{abc}{}_d) X^I_c \ .
\end{equation}
The first term is simply a translation, as expected. Just as before the second term must be interpreted as a symmetry:
\begin{equation}
\delta X^I_d  = \tilde \Lambda^c{}_dX^I_d\;,\qquad \tilde \Lambda^c{}_d = 2i\bar\epsilon_2\Gamma^{JK}\epsilon_1 X^J_aX^K_bf^{abc}{}_d\ .
\end{equation}
Indeed this is be a gauge symmetry since $\tilde \Lambda^c{}_d$ depends on $X^J_b$ which in turn depends on $x^\mu$.

Next we must introduce a gauge field for this gauge symmetry. Following standard techniques we define
\begin{equation}
D_\mu X^I_d = \partial_\mu X^I_d - \tilde A_\mu^c{}_d X_c^I\ ,
\end{equation}
and similarly for $\Psi_d$. This is gauge covariant provided that 
\begin{equation}
\delta \tilde A_\mu^c{}_d = \partial_\mu \tilde \Lambda^c{}_d + \tilde A_\mu^c{}_e \tilde \Lambda^e{}_d - \tilde \Lambda^c{}_e \tilde A_\mu^e{}_d\ .
\end{equation}
We can also compute the field strength from $[D_\mu,D_\nu]X^I_b = \tilde F_{\mu\nu}{}^a{}_bX^I_a$ and find
\begin{equation}
 \tilde F_{\mu\nu}{}^a{}_b = \partial_\nu \tilde A_\mu{}^a{}_b -  \partial_\mu \tilde A_\nu{}^a{}_b - \tilde A_{\mu}^a{}_c\tilde A_\nu{}^c{}_b +  \tilde A_{\nu}^a{}_c\tilde A_\mu{}^c{}_b\;.
\end{equation}
These are familiar expression from gauge theory. 

We are now in a position to postulate the complete set of supersymmetry transformations:
 \begin{eqnarray}\label{susygauged2}
\nonumber \delta X^I_a &=& i\bar\epsilon\Gamma^I\Psi_a\\
\delta \Psi_a &=& D_\mu X^I_a\Gamma^\mu \Gamma^I\epsilon -\frac{1}{3!}
X^I_bX^J_cX^K_d f^{bcd}{}_{a}\Gamma^{IJK}\epsilon \\
\nonumber \delta\tilde A_{\mu}{}^b{}_a &=& i\bar\epsilon
\Gamma_\mu\Gamma_IX^I_c\Psi_d f^{cdb}{}_{a}.
\end{eqnarray}
These supersymmetries close on-shell into translations and
gauge transformations. In the process one discovers that   the structure constants $f^{abc}{}_d$ must obey:
\begin{equation}\label{FIa}
f^{abc}{}_gf^{efg}{}_d = f^{efa}{}_gf^{gbc}{}_d + f^{agc}{}_df^{efb}{}_g + f^{abg}{}_d f^{efc}{}_g\ ,
\end{equation}
which is known as the fundamental identity.

As we mentioned above, on-shell closure means that we can deduce the equations of motion directly from the algebra. In this case we find
\begin{eqnarray}\label{EOMS}
\nonumber\Gamma^\mu D_\mu\Psi_a
+\frac{1}{2}\Gamma_{IJ}X^I_cX^J_d\Psi_bf^{cdb}{}_{a}&=&0\\
 D^2X^I_a-\frac{i}{2}\bar\Psi_c\Gamma^I_{\ J}X^J_d\Psi_b f^{cdb}{}_a
   -\frac{\partial V}{\partial X^{Ia}}    &=& 0 \\
\nonumber \tilde F_{\mu\nu}{}^b{}_a
  +\varepsilon_{\mu\nu\lambda}(X^J_cD^\lambda X^J_d
+\frac{i}{2}\bar\Psi_c\Gamma^\lambda\Psi_d )f^{cdb}{}_{a}  &=& 0.
\end{eqnarray}

To construct a lagrangian we need to introduce and inner-product on the 3-algebra: $\langle X^I,X^J\rangle = h^{ab}X^I_aX^J_b$. Invariance of the inner product under 
$\delta X^I_a=\tilde\Lambda^b{}_aX^I_b$ implies that
\begin{equation}
f^{abcd}=h^{de}f^{abc}{}_e=f^{[abcd]}\ ,
\end{equation}
which is analogous to the similar condition we obtain before for D-branes. The  lagrangian can now be written as
\begin{eqnarray}\label{action}
\nonumber {\cal L} = -\frac{1}{2}D_\mu X^{aI}D^\mu X^{I}_{a}
+\frac{i}{2}\bar\Psi^a\Gamma^\mu D_\mu \Psi_a 
+\frac{i}{4}\bar\Psi_b\Gamma_{IJ}X^I_cX^J_d\Psi_a f^{abcd}- V +{\cal L}_{CS}\ ,
\end{eqnarray}
where 
\begin{eqnarray}
V= \frac{1}{12}X^I_aX^J_bX^K_cX^I_eX^J_fX^K_g f^{abcd}f^{efg}{}_d\ ,
\end{eqnarray}
is the potential and
\begin{equation}
{\cal L}_{CS}= \frac{1}{2}\varepsilon^{\mu\nu\lambda}(f^{abcd}A_{\mu
ab}\partial_\nu A_{\lambda cd} +\frac{2}{3}f^{cda}{}_gf^{efgb}
A_{\mu ab}A_{\nu cd}A_{\lambda ef}).
\end{equation}
is a `twisted' Chern-Simons term. Indeed an earlier attempt at describing M2-branes with Chern-Simons-matter theories was made in \cite{Schwarz:2004yj}.
Note that ${\cal L}_{CS}$ is written in
terms of $A_{\mu ab}$ and not the physical field $\tilde
A_{\mu}{}^b{}_a = A_{\mu cd}f^{cdb}{}_a$ that appears in the
supersymmetry transformations and equations of motion. However,
one can check that ${\cal L}_{CS}$ is invariant under shifts of
$A_{\mu ab}$ that leave $\tilde A_{\mu}{}^b{}_a$ invariant. Thus
it is locally well defined as a function of $\tilde A_{\mu}{}^b{}_a$.

This theory is invariant under 16
supersymmetries and an $SO(8)$ R-symmetry.  It is also scale invariant (and in fact fully conformally
invariant \cite{Bandres:2008vf}). These are all the continuous
symmetries that are expected of multiple M2-branes.  It is weakly coupled in the limit that $f^{abc}_{d}\to 0$ and thus apparently has a continuous deformation parameter: $f^{abc}_{d}\to \lambda f^{abc}_{d}$. However we will see that due to the Chern-Simons term, $\lambda$ is discrete. Thus there are no continuous free parameters, as expected from M-Theory. For a recent discussion of quantum Chern-Simons theories see \cite{Gaiotto:2007qi}.

Note that the
Chern-Simons term naively breaks the parity that is expected to be a
symmetry of the M2-brane worldvolume. However, we can make the
lagrangian parity invariant if we assign an odd parity to $f^{abcd}$.  In
particular, if we invert $x^2 \to -x^2$, we must then require
that  $X^I_a$ and $\tilde A^{\ a}_{\mu\ b}$ be parity even for $\mu =
0,1$; $ \tilde A^{\ a}_{2\ b}$ and $f^{abcd}$ be parity odd; and
$\Psi_a\to \Gamma_2\Psi_a$.  Note that this assignment implies that
$A_{\mu ab}$ is parity odd for $\mu=0,1$, while $A_{2 ab}$ is
parity even.

This seems like complete success: we have got all that we asked for. Unfortunately we will see that there is essentially just one solution to the fundamental identity and therefore the set of such lagrangian field theories is too small to apply to an arbitrary number of M2-branes.

\subsection{3-algebras}

We have seen that the construction of M-Theory leads to the notion of a 3-algebra. That is to say the structure constants $f^{abc}{}_d$ define a triple product on the algebra:
\begin{equation}
[T^a,T^b,T^c] = f^{abc}{}_d T^d\ .
\end{equation}
In this case the triple product is totally antisymmetric and is sometimes called a Lie 3-algebra. However in the next section we will see that this assumption needs to be dropped in general. Note that the appearance of a 3-algebra has been forced on us by the symmetries of the problem - most notably conformal symmetry which requires a cubic term in $\delta \Psi_a$.

The fundamental identity eq.(\ref{FIa}) ensures that the set of all $\tilde\Lambda^a{}_d = \Lambda_{bc}f^{abc}{}_d$ for some $\Lambda_{bc}$ forms a closed set under matrix commutation. Thus the 3-algebra defines a Lie algebra  generated by the elements $\tilde \Lambda^a{}_d$, along with a natural representation which acts on the 3-algebra itself. So the underlying gauge symmetry of the theory is an ordinary  gauge theory based on Lie algebras and the vector space of the 3-algebra plays the
role of a preferred representation.

Let us consider the analogy with Lie algebras and the adjoint map more carefully. In particular we define the map $ad_{A,B}(X) = [X, A, B]$ and require that this is a derivation:
\begin{eqnarray}
ad_{A,B}( [X,Y,Z] )= [ ad_{A,B}(X), Y, Z] + [X,ad_{A,B}( Y),Z]+[X,Y,ad_{A,B}( Z)]\ .
\end{eqnarray}
This further requires that the triple product satisfies the so-called fundamental identity:
\begin{equation}\label{FI}
[[X,Y,Z], A,B] = [[X,A,B],Y,Z] + [X,[Y,A,B],Z] + [X,Y,[Z,A,B]]\ ,
\end{equation}
which is equivalent in eq.(\ref{FIa}). If we relax the assumption that the triple product is totally anti-symmetric then there is a one-to-one relation between a 3-algebras and the set of Lie algebras along with a specified representation\footnote{Technically one also requires that all these spaces, the 3-algebra, the Lie algebra and the representation space, all have invariant inner-products}. The key observation is that, given a representation with generators $(T^r)^{a}{}_b$, $r = 1,2,...,{\rm rank}(G)$ $a,b = 1,2,...,{\rm dim\ Rep}(G)$ then there is a natural triple product with structure constants:
\be
f^{abc}{}_d = \sum_r \kappa_{rs}h^{be}(T^r)^{a}{}_e(T^s)^{c}{}_d\ ,
\ee
where $\kappa_{rs}$ and $h^{ab}$ are invariant metrics of the algebra and representation respectively. Conversely given $f^{abc}{}_d$ we can reconstruct the Lie-bracket and representation of the Lie algebra $G$ by looking at the action of the $ad_{A,B}$ maps.

Thus the 3-algebra is a convenient way of packaging up all the information of the gauge theory (gauge algebra, matter representation and couplings). Asking for various symmetry properties of the triple product then selects out certain pairs of Lie algebras and representations as special. Thus the triple product naturally determines the amount of supersymmetry and (somewhat unusual) gauge group. In contrast to Yang-Mills theories where any gauge group is sufficient. The interested reader should consult \cite{Faulkner,Filipov,Jacobson, Kasymov,Lister, Takhtajan:1993vr} for some mathematical discussion of 3-algebras and  \cite{deMedeiros:2008bf,deMedeiros:2009hf,MendezEscobar:2010zq,deAzcarraga:2010mr} for physical applications  to M2-branes and other systems,  as well as the references therein.

We also note that, despite first appearances, models made from 3-algebras with indefinite metrics can make physical sense. In particular the negative normed states can be gauged away in a similar way to the timelike components of gauge fields. The theories that results from these models ultimately reduce to the Yang-Mills gauge theories of D$p$-branes \cite{Gomis:2008uv,Benvenuti:2008bt,Ho:2008ei,DeMedeiros:2008zm,Bandres:2008kj,Ezhuthachan:2008ch,Gomis:2008be}. In addition there are positive 3-algebras with infinite dimension. These are constructed from the famous Nambu bracket \cite{Nambu:1973qe}. Models based on these algebras have also appeared in the literature where they have been related to M5-branes \cite{Ho:2008nn,Bandos:2008fr,Pasti:2009xc,Park:2008qe,Gustavsson:2010nc}.

\subsection{An Example: BLG}

Let us look more carefully at the case at hand where the triple product is totally anti-symmetric and the metric is positive definite. If we assume that the metric $h^{ab}$ is positive definite, so that the kinetic energies and potential are all positive, then it turns out that there is essentially a unique choice for $f^{abcd}$ \cite{nagy-2007,Papadopoulos:2008sk,Gauntlett:2008uf}:
\be
f^{abcd} = \frac{2\pi}{k}\varepsilon^{abcd}\ ,
\ee
where the 3-algebra has dimension four; $a,b,...=1,2,3,4$. The gauge algebra is generated by $\tilde \Lambda^a{}_b$ is simply the space of all anti-symmetric $4\times 4$ matrices. This is of course $so(4)=su(2)\oplus su(2)$. This split is realized by noting that the self-dual and anti-self dual parts of $\tilde \Lambda{}^a{}_b$ commute with each other. Thus we write 
\begin{equation}
\tilde A_\mu{}^a{}_b = \tilde A^+_\mu{}^a{}_b+\tilde A^-_\mu{}^a{}_b\ ,
\end{equation}
where $\tilde A^\pm_\mu{}^a{}_b$ is the (anti)-self-dual part of $\tilde A_\mu{}^a{}_b$. In this case the `twisted' Chern-Simons term can be written as
\bea
{\cal L}_{CS} &=& \frac{k}{8\pi}\epsilon^{\mu\nu\lambda} (\tilde A^+_\mu{}^a{}_b \partial_\nu A^+_{\lambda}{}^b{}_a
+\frac{2}{3}A^+_{\mu}{}^a{}_bA^+_{\nu}{}^b{}_cA^+_{\lambda}{}^c{}_a )\\ 
\nonumber &&-
\frac{k}{8\pi}\epsilon^{\mu\nu\lambda}( \tilde A^-_\mu{}^a{}_b \partial_\nu A^-_{\lambda}{}^b{}_a
+\frac{2}{3}A^-_{\mu}{}^a{}_bA^-_{\nu}{}^b{}_cA^-_{\lambda}{}^c{}_a )\ .
\eea
The action of parity changes the sign of each of the two terms ${\cal L}_{CS}$ and in addition  flipping the sign of $f^{abcd}$ can now be seen as swapping the two $su(2)$ subalgebras. This leads to the lagrangian first constructed in \cite{Bagger:2007jr}. The action can then be reformulated in more familiar language as a Chern-Simons-Matter theory with gauge algebra $su(2)\times su(2)$ and matter fields in the bi-fundamental \cite{VanRaamsdonk:2008ft}.

Next we must examine the quantization condition on $k$. Let us first consider a single $su(N)$ gauge field $A_\mu$ and Chern-Simons term
\begin{equation}
{\cal L}_{su(2)} =  \frac{k}{4\pi}\epsilon^{\mu\nu\lambda}{\rm tr}(A_\mu\partial_\nu A_\lambda - \frac{2i}{3}A_\mu A_\nu A_\lambda)\ ,
\end{equation}
where ${\rm tr}$ is the trace in the fundamental (\ie\ $N\times N$) representation
Under a large gauge transformation one has \cite{Deser:1981wh}
\begin{equation}
\int d^3 x{\cal L}_{su(N)} \to \int d^3 x{\cal L}_{su(N)}  + {2\pi k w}\ ,
\end{equation}
where $w\in \mathbb Z$ is the winding number of the gauge transformation. For the quantum theory to be well defined we require that $e^{i\int {\cal L}_{su(N)}}$ is invariant. This fixes $k\in \mathbb Z$. However in our case, where $N=2$, the coefficient is $\frac{k}{8\pi}$ not $\frac{k}{4\pi}$. On the other hand the gauge fields $\tilde A^\pm_\mu{}^a{}_b$ are acting in the $4\times 4$  and not the $2\times 2$ representation and this gives us an extra factor of $2$. As a result we also find $k \in \mathbb Z$.
 
Thus we have obtained a family of lagrangians with 16 supersymmetries, $SO(8)$ R-symmetry and conformal invariance. There are no free parameters but there is a discrete parameter $k\in \mathbb Z$ and the theory is weakly coupled in the limit $k\to\infty$. We would also like to mention an additional choice. Namely the lagrangian is determined entirely in terms of the 3-algebra structure constants. This in turn determines the local gauge structure however there still remains the choice of the global gauge group. In the case at hand we can either take this to be $SU(2)\times SU(2) = Spin(4)$ or $(SU(2)\times SU(2))/{\mathbb Z}_2 = SO(4)$. Therefore there are actually two theories, each of which is specified by a quantized coupling constant $k\in \mathbb Z$. This subtlety did not arise in the D$p$-brane case as the matter fields there (in the maximally supersymmetric case, not for less supersymmetric theories) are in the adjoint and this is blind to such global properties.

\section{M2-branes and Chern-Simons-Matter Theories: 12 Supersymmetries}
\label{abjmetc}

The success of the previous construction calls out for some sort of generalization. From the perspective that we have presented we need to look for less restrictive conditions on the 3-algebra. This in turn means that we need to look for less supersymmetry. Unlike higher dimensions, it turns our that the next smallest amount of supersymmetry in three-dimensions consists of 12 supersymmetries. This will lead to the celebrated ABJM models which we now discuss.

If we reduce the number of supersymmetries  from 16 (${\cal N}=8$) to 12 (${\cal N}=6$)\footnote{Here ${\cal N}$ counts the number of spinors rather than the total number of spinor components. Since each three-dimensional real spinor has  two components  a single spinor ${\cal N}=1$ would generate two supersymmetries.} then we will also reduce the $SO(8)$ R-symmetry to $SU(4)\cong SO(6)$.  There will also be a left-over $U(1)$ symmetry under which the ${\cal N}=6$ supersymmetries are neutral (but the missing ${\cal N}=2 $ supersymmetries are charged under).
This $U(1)$ symmetry will be important and  will ultimately be  gauged.  

To conveniently describe all these symmetries introduce four
complex scalar fields $Z^A_a$, $A=1,2,3,4$, as well
as their complex conjugates $\bZ_{A}^a$.  Similarly, we denote the
fermions by $\psi_{Aa}$ and their complex conjugates by $\psi^{Aa}$. These are now complex 2 component spinors.
A raised $A$ index indicates that the field
is in the $\bf 4$ of $SU(4)$; a lowered index transforms in the
$\bar{\bf 4}$. We assign $Z^A_a$ and $\psi_{Aa}$ a $U(1)$ charge of
1. Complex conjugation raises or lowers the $A$ and $a$ indices and flips
the sign of the $U(1)$ charge.
The supersymmetry generators $\epsilon_{AB}$ are in the $\bf 6$ of
$SU(4)$ with vanishing $U(1)$ charge.  They satisfy the reality
condition $\epsilon^{AB} = \frac{1}{2}
\varepsilon^{ABCD}\epsilon_{CD}$.

One can follow the same discussion that we gave above to arrive at the following form for the supersymmetry algebra
 that preserve the $SU(4)$, $U(1)$ and conformal symmetries. We will not go through the full derivation here, for that see \cite{Bagger:2008se}.
 The result is that the most general transformations are
\begin{eqnarray}\label{finalsusy}
\nonumber  \delta Z^A_d &=& i\bar\epsilon^{AB}\psi_{Bd} \\
\nonumber  \delta \psi_{Bd} &=& \gamma^\mu D_\mu Z^A_d\epsilon_{AB} +
  f^{ab}{}_{cd}Z^C_a Z^A_b \bZ_{C}^c \epsilon_{AB}+
  f^{ab}{}_{cd} Z^C_a Z^D_{b} \bZ_{B}^c\epsilon_{CD} \\
  \delta \tilde A_\mu{}^c{}_d &=&
-i\bar\epsilon_{AB}\gamma_\mu Z^A_a\psi^{Bb} f^{ca}{}_{bd} +
i\bar\epsilon^{AB}\gamma_\mu \bZ_{A}^b\psi_{Ba} f^{ca}{}_{bd} .
\end{eqnarray}
These close into translations and gauge transformations:
\begin{equation}\label{symgen}
v^\mu = \frac{i}{2}\bar\epsilon_2^{CD}\gamma^\mu\epsilon_{1CD},\qquad
\Lambda^c{}_{b}=i(\bar\epsilon^{DE}_2\epsilon_{1CE}-
\bar\epsilon^{DE}_1\epsilon_{2CE})\bZ_{D}^cZ^C_b.\\
\end{equation}
provided that the fields satisfy the on-shell conditions:
\begin{eqnarray}
 \gamma^\mu D_\mu\psi_{Cd} &=& f^{ab}{}_{cd} \psi_{Ca}
Z^D_b\bZ_{D}^c+2f^{ab}{}_{cd}\psi_{Da}Z^D_b\bZ_{C}^c-\varepsilon_{CDEF}f^{ab}{}_{cd}\psi^{Dc} Z^E_aZ^F_b.\\
\tilde F_{\mu\nu}{}^c{}_d &=&-
\varepsilon_{\mu\nu\lambda}\left(D^\lambda Z^A_a \bZ_{A}^b-
Z^A_aD^\lambda \bZ_{A}^b
-i\bar\psi^{Ab}\gamma^\lambda\psi_{Aa}\right)f^{ca}{}_{bd},
\end{eqnarray}

Furthermore the structure constants $f^{ab}{}_{cd} = -f^{ba}{}_{cd}$ define a triple product
\begin{equation}
[T^a,T^b;T_c] = f^{ab}{}_{cd}T^d\ ,
\end{equation}
which must satisfy the following fundamental identity:
\begin{equation}\label{complexFI}
f^{ef}{}_{gb}f^{cb}{}_{ad} +f^{fe}{}_{ab}f^{cb}{}_{gd}+
f^*_{ga}{}^{ fb} f^{ce}{}_{bd}+f^*_{ag}{}^{ eb} f^{cf}{}_{bd}=0.
\end{equation}
We also see that in this case the triple product is linear and anti-symmetric in its first two entries but complex anti-linear in the third. The role of this fundamental identity is just the same as in the real case above. In particular it implies that $ad_{A,B}(X)  = [X,A;\bar B]$ acts as a derivation and generates a Lie algebra action on the fields.

Let us now construct an invariant lagrangian.  We have seen that
the supersymmetry algebra closes into a translation plus a gauge
transformation.  On the field $\bZ_{A}^d$, we find
\begin{equation}
[\delta_1,\delta_2]\bZ_{A}^d = v^\mu D_\mu \bZ_{A}^d +
\Lambda^{*b}{}_{c }f_{ab}{}^ {cd}\bZ_{A}^a,
\label{cl}
\end{equation}
with $v$ and $\Lambda^c{}_{ b}$ given in eq.(\ref{symgen}).  The second
term is a gauge transformation, $\delta_\Lambda \bZ_{Ad^d} =
\Lambda^{*b}_{c } f^*_{ab}{}_{cd}\bZ_{A}^a =
-\Lambda^b_{ c} f^*_{ab}{}^{cd} \bZ_{A}^a$.  In this case, to construct an lagrangian 
 we need the metric to be gauge invariant, namely
$\delta_\Lambda (h_a{}^{b} \bZ_{A}^a Z^A_b)=0$.  Therefore we
must require
\begin{equation}\label{last}
f^{ab}{}_{cd} = {f^*}_{cd}{}^{ab},
\end{equation}
where $f^{ab}{}_{cd} =f^{ab}{}_{ce} h^e{}_{ d}$.  This implies that
$(\tilde\Lambda^{c }{}_d)^* = -\tilde \Lambda^{d}{}_c$, where
\begin{equation}
\tilde \Lambda^{c}{}_d = \Lambda^b{}_{ a}f^{ca}{}{bd},
\end{equation}
so the transformation parameters $\tilde \Lambda^c{}_d$
are unitary matrices. Furthermore the fundamental identity ensures that they are a Lie subalgebra, \ie\ closed under ordinary matrix commutation.

With these results, it is not hard to show that an invariant
lagrangian (up to boundary terms) is given by
\begin{eqnarray}\label{lagrangian}
\nonumber {\cal L} &=& - D^\mu \bZ_A^a D_\mu Z^A_a -
i\bar\psi^{Aa}\gamma^\mu D_\mu\psi_{Aa} -V+{\cal L}_{CS}\\[2mm]
&& -i f^{ab}{}_{cd}\bar\psi^{Ad} \psi_{Aa}
Z^B_b\bZ_{B}^c+2if^{ab}{cd}\bar\psi^{Ad}\psi_{Ba}Z^B_b\bZ_{A}^c\\
\nonumber
&&+\frac{i}{2}\varepsilon_{ABCD}f^{ab}{}_{cd}\bar\psi^{Ad}\psi^{Bc} Z^C_aZ^D_b
-\frac{i}{2}\varepsilon^{ABCD}f^{cd}{}_{ab}\bar\psi_{Ac}\psi_{Bd}\bZ_{C}^a\bZ_{D}^b\
,
\end{eqnarray}
where the potential is
\begin{equation}
V = \frac{2}{3}\Upsilon^{CD}_{Bd}\bar\Upsilon_{CD}^{Bd} ,
\end{equation}
with
\begin{equation}
\Upsilon^{CD}_{Bd} = f^{ab}_{cd}Z^C_aZ^D_b\bZ_{B}^c
-\frac{1}{2}\delta^C_Bf^{ab}{}_{cd}Z^E_aZ^D_b\bZ_{E}^c+\frac{1}{2}\delta^D_Bf^{ab}{}_{cd}Z^E_aZ^C_b\bZ_{E}^c.
\end{equation}

The `twisted' Chern-Simons term ${\cal L}_{CS}$ is given by
\begin{equation}
{\cal
L}_{CS}=\frac{1}{2}\varepsilon^{\mu\nu\lambda}\left(f^{ab}{}_{cd}A_{\mu}^c{}_b\partial_\nu A_{\lambda}^b{}_a +\frac{2}{3}f^{ac}{}_{dg}f^{ge}{}_{fb}
A_{\mu}{}^b{}_a A_{\nu}{}^d{}_cA_{\lambda}{}^f{}_e\right).
\end{equation}
It satisfies
\begin{equation}
\frac{\delta{\cal L}_{CS}}{\delta \tilde A_{\lambda}{}^{a}{}_b}f^{ac}{}_{db} =
\frac{1}{2}\varepsilon^{\lambda\mu\nu}\tilde F_{\mu\nu}{}^{c}{}_d,
\end{equation}
up to integration by parts,
where $\tilde F_{\mu\nu}{}^a{}_b = -\partial_\mu \tilde
A_\nu{}^a{}_b+\partial_\nu \tilde A_\mu{}^a{}_b +  \tilde
A_\nu{}^a{}_e\tilde A_\mu{}^e{}_b- \tilde A_\mu{}^a{}_e\tilde
A_\nu{}^e{}_b $.  Just as before, this term can be
viewed as a function of $\tilde A_\mu{}^c{}_d$ and not $A_{\mu c}{}^d$.

Just as with the maximally supersymmetric case one also can add massive deformations that are consistent with all 12 supersymmetries \cite{Hosomichi:2008qk}, again 
corresponding to turning on background supergravity flux \cite{Lambert:2009qw}.

\subsection{Examples: ABJM and ABJ}

Finally we wish to give an example of a suitable triple product. As promised we will be able to obtain infinitely many theories. Most importantly one finds the gauge groups $U(N)\times U(M)$ for any $N$ and $M$ and constructs the famous ABJM models when $N=M$ \cite{Aharony:2008ugx} and ABJ models, for $N\ne M$,  \cite{Aharony:2008gk}. The lagrangians for these theories were first obtained in \cite{Benna:2008zy,Hosomichi:2008jb}. In addition there are other possible gauge groups and a classification was provided in \cite{Schnabl:2008wj}.

To continue let $Z^A$ be $N\times M$ complex matrices and define
\begin{equation}\label{6triplep}
[Z^A,Z^B;\bZ_C] = \frac{2\pi}{k}(Z^AZ^\dag_C Z^B - Z^BZ^\dag_C Z^A)\ .
\end{equation}
One can show that this satisfies eq.(\ref{complexFI}).
What is the Lie algebra generated by this triple product? The transformations $\delta Z^A_a = \Lambda^c{}_{d}f^{ab}{}_{cd}Z^A_b$ take the form, assuming matrix multiplication,
\begin{equation}
\delta Z^A  = \Lambda^LZ^A-Z^A\Lambda^R\ .
\end{equation}
Here $\Lambda^L$ and $\Lambda^R$ are $N\times N $ and $M\times M$ matrices respectively. Thus we see that the Lie algebra is $u(N)\oplus u(M)$ and the fields are in the bi-fundamental representation. Thus the action of gauge fields is 
\be
D_\mu Z^A = \partial_\mu Z^A - iA^L_\mu Z^A+iZ^AA_\mu^R\ ,
\ee 
where $A^L_\mu$ and $A^R_\mu$ are the $N\times N$ and $M\times M$ complex matrix valued gauge fields of $u(N)\oplus u(M)$. Note also that the $U(1)$ symmetry that rotates the phase of $Z^A$ and which we obtained from the breaking of $SO(8)$ R-symmetry to $SO(6)\times U(1)$ has become gauged\footnote{In fact,  since $u(N)=u(1)\oplus su(N)$ there are two abelian $u(1)$ factors in $u(N)\oplus u(M)$. It turns out that only the relative $u(1)$  couples to the matter fields (through phase rotations).}. In addition, just as in the case above, the Chern-Simons term can be written as
\bea
{\cal L}_{CS} &=& \frac{k}{4\pi}\epsilon^{\mu\nu\lambda}{\rm tr} (  A^L_\mu{}  \partial_\nu A^L_{\lambda}{} 
-\frac{2i}{3}A^L_{\mu}{} A^L_{\nu} A^L_{\lambda}  )\\ 
\nonumber &&-
\frac{k}{4\pi}\epsilon^{\mu\nu\lambda}{\rm tr} (   A^R_\mu  \partial_\nu A^R_{\lambda} 
-\frac{2i}{3}A^R_{\mu} A^R_{\nu} A^R_{\lambda}  )\ ,
\eea
so that again we must take $k\in {\mathbb Z}$.

Actually there is a slight subtlety here. In the ABJM case where $N=M$ one finds that the action of the two $U(1)$ subgroups coming from $U(n)$ and $U(m)$ cancel since
 ${\rm tr}(\Lambda^L)={\rm tr}(\Lambda^R)$. So in this case, following the construction of the theory through 3-algebras that we considered here,  we are left with simply $su(N)\oplus su(N)$ Chern-Simons theory.
To fix this we can put the $u(1)$ back by gauging the global $U(1)$ by hand. This was discussed in some detail in \cite{Lambert:2010ji}. We do not need to go into the details here. The effect is to simply include the missing components of $A^L_\mu$ and $A^R_\mu$ that are proportional to the identity matrix into the action to get the complete $U(N)\times U(N)$ theory. This can be done preserving all the supersymmetries. 

To summarize we have found several classes of theories, all with a discrete parameter $k\in \mathbb Z$ so that they become weakly coupled as $k\to \infty$. The Lie algebras that we have found here are $su(N)\oplus su(M)\oplus u(1)$ and $su(N)\oplus su(N)$. In fact there is another choice (not given by eq.(\ref{6triplep})) which leads to an $sp(2N)\oplus so(2)$ lagrangian \cite{Chen:2009cwa,Bagger:2010zq,Palmkvist:2011aw}. This list agrees with the analysis of \cite{Schnabl:2008wj}.  These lagrangians can lead to slightly different physical theories corresponding to the different possible global forms of the gauge group. But of most interest for us here are the ABJM and ABJ $U(N)\times U(M)$ theories which are most readily relevant for M2-branes.

\subsection{Spacetime Interpretation}

Just as in the case of D$p$-branes we'd like to determine the spacetime interpretation and hence determine the vacuum moduli space. For these Chern-Simons-Matter theories of $N$ M2-branes this was first performed in \cite{Lambert:2008et,VanRaamsdonk:2008ft,Aharony:2008ugx}.  We'd also like to learn the meaning of the parameter $k\in \mathbb Z$. For $N$ M2-branes in  ${\mathbb R}^{1,10}$ the moduli space should be
\be\label{Mguess}
{\cal M} = \frac{({\mathbb C}^{4})^N}{S_N}\ .
\ee
To begin we must decide what the vacuum solutions are. These need to preserve all the supersymmetries and, an inspection of eq.(\ref{finalsusy}), reveals that we therefore require that $[Z^A,Z^B;\bZ_C]=0$;
\be
Z^AZ^\dag_C Z^B = Z^BZ^\dag_C Z^A\ ,
\ee
for all $A,B,C$. Let us assume that $M=N+l\ge N$ and think of $Z^A$ in terms of the block form
\be
Z^A = \left(\begin{array}{ll}W^A_{N\times N}& V^A_{l\times N}\end{array}\right)\ .
\ee
We then find that, for generic vacua, the $W^A$ components should all commute with each other and the $V^A$ components should vanish. By taking gauge transformations of the form $g_L=g_R\in U(N)\subset U(M)$ the action of $U(N)\times U(M)$ is reduced to the adjoint action of $U(N)$ when acting on $W^A$. This allows us to diagonalize:
\be
Z^A = \left(\begin{array}{ll}\begin{array}{lll}z^A_1 &&\\ &\ddots &\\ && z_N^A\\ \end{array}& \begin{array}{l}0 \\ \vdots \\ 0\\ \end{array}\end{array}\right)\ .
\ee 
In addition, just as was the case with D$p$-branes, we can also take $g_L=g_R$ of the form eq.(\ref{U2}) to induce the identification  $z^A_i\leftrightarrow z^A_j$ for any pair $i,j=1,..,N$. Therefore we can associate $z^A_i$ with the position of the $i$th M2-brane in the transverse space ${\mathbb C}^4$. Further identification by gauge transformations then says that these are indistinguishable objects and we arrive at eq.(\ref{Mguess}).

However there are important remaining gauge transformations to consider that did not arise in the case of D$p$-branes. In particular we can take $g_R=1$ and 
\be
g_L = \left(\begin{array}{lll}e^{i\theta_1} &&\\ &\ddots &\\ && e^{i\theta_N}\end{array}\right)\ .
\ee
This acts as a phase rotation on the $z^A_i$, $i=1,...,N$. This is not the end of the story though since there also also gauge fields that preserve the vacuum moduli space, in particular we can turn on gauge fields of the form
\be
A_\mu^L= \left(\begin{array}{lll}a_\mu^1 &&\\ &\ddots &\\ && a_\mu^N\\ \end{array}\right)\ , \qquad A_\mu^R= \left(\begin{array}{lllll}b_\mu^1 &&&&\\ &\ddots &&&\\ && b_\mu^N&&\\&&&0&\\&&&&\ddots \end{array}\right)\ . 
\ee 
The effective lagrangian for the vacuum fields is
\be
{\cal L}_{vacuum} = -\sum^N_{i=1}D_\mu z^A_i D^\mu \bar z_{Ai} + \frac{k}{4\pi}\sum^N_{i=1}\varepsilon^{\mu\nu\lambda} B^i_\mu\partial_\nu A^i_\lambda\ ,
\ee
where $D_\mu z^A_i = \partial_\mu z_i^A - iA_\mu^i z^A_i$, $A_\mu^i  =a_\mu^{i}-b^i_\mu $ and $B_\mu^i = a_\mu^{i}+b^i_\mu $.
Note that $B^i_\mu$ only appears in the Chern-Simons term. We can therefore integrate it out and deduce that
\be\label{Aflat}
A^i_\lambda = \frac{1}{k}\partial_\lambda \sigma_i\ ,
\ee
for some scalars $\sigma_i$. In this case the lagrangian reduces simply to
\be
{\cal L}_{vacuum} = -\sum^N_{i=1}\partial_\mu w^A_i \partial^\mu \bar w_{Ai}\ ,
\ee
where $w^A_i = e^{i\sigma_i/k}z^A_i$.

Our next step is to notice that $\sigma$ is periodic. To see this we go back a step and write the Chern-Simons term in ${\cal L}_{vacuum}$ as, using integration by parts, 
\bea
\frac{k}{4\pi}\sum^N_{i=1}\varepsilon^{\mu\nu\lambda} B^i_\mu\partial_\nu A^i_\lambda &=& - \frac{1}{4\pi}\sum^N_{i=1}\varepsilon^{\mu\nu\lambda} \partial_\nu B^i_\mu \partial_\lambda\sigma_i  \\
&=&
\frac{1}{8\pi}\sum^N_{i=1}\varepsilon^{\mu\nu\lambda}  (F^{Li}_{\mu\nu}+F^{Ri}_{\mu\nu}) \partial_\lambda\sigma_i \nonumber\\ 
&=&-
\frac{1}{4\pi}\sum^N_{i=1}\varepsilon^{\mu\nu\lambda}  \partial_\lambda F^{Li}_{\mu\nu}  \sigma_i\ , \nonumber
\eea
where in the last line we have the condition eq.(\ref{Aflat}) to set $F^{Li}_{\mu\nu}=F^{Ri}_{\mu\nu}$. We now observe that, by the standard Dirac  quantization rule for magnetic fluxes
\be\label{Dirac}
\frac{1}{4\pi}\int d^3 x \varepsilon^{\mu\nu\lambda}  \partial_\lambda F^{Li}_{\mu\nu} = \frac{1}{2\pi} \int dF^{Li} \in {\mathbb Z}\ .
\ee
This means that the action is invariant under shifts $\sigma_i\to \sigma_i+2\pi$. It then follows that $w^A_i\cong e^{2\pi i/k}w^A_i$. In this way we see that the true moduli space is
\be\label{M2vac}
{\cal M} = \frac{({\mathbb C}^{4}/{\mathbb Z}_k)^N}{S_N}\ .
\ee
Thus we see that the M2-branes are actually moving in an ${\mathbb C}^{4}/{\mathbb Z}_k$ transverse space. Therefore, to describe $N$ M2-branes in ${\mathbb R}^8$ we must take $k=1$. In this case the lagrangian is strongly coupled, as expected for the M2-brane worldvolume theory. 

We also see how the limit $k\to \infty$ produces weak coupling. To see this we can write ${\mathbb R}^8$ as a cone\footnote{This is a fancy way of saying use `spherical' coordinates: $ds^2_{{\mathbb R}^8} = dr^2 +r^2 ds^2_{S^7}$.} over $S^7$ and then view  $S^7$ as fibration of $S^1$ over $S^6$: $S^7= S^6 \ltimes S^1$. Intuitively, as $k \to\infty$, the action of ${\mathbb Z}_k$ on $S^1$ amounts to shrinking it to zero size. We are therefore driven into the regime of weakly coupled type IIA String Theory and the M2-branes should be described by D2-branes and hence super-Yang-Mills. 

This can be made more precise by the so-called novel Higg's mechanism of \cite{Mukhi:2008ux}. Here one gives a large vev to one of the scalar fields, say, 
\be\label{vevmp}
\langle Z^4\rangle  = iv 1_{N\times N} \ ,
\ee
corresponding to putting all the M2-branes at some distance $v$ away\footnote{More precisely, since $v^2$ has dimensions of mass, the physical distance is $v/\sqrt{T_{M2}}$ where $T_{M2}$ is the tension of an M2-brane.} from the origin along the ${\rm Im} Z^4$ direction (for simplicity we just consider the $N=M$ ABJM case).  This breaks the gauge group from $U(N)\times U(N)\to U(N)$. Furthermore for large $v$ one is far from the origin and for small fluctuations ${\mathbb R}^8$ looks like ${\mathbb R}\times S^7$. One finds that, as a consequence of the Chern-Simons terms,  the remaining gauge field $A^L_\mu+A^R_\mu$ becomes dynamical by `eating' the scalar field ${\rm Im} Z^4$ (recall that in a standard Higg's mechanism a massless gauge field becomes massive by eating a scalar - here a non-dynamical gauge field becomes dynamical by eating a scalar). The resulting low energy theory therefore has a dynamical $u(N)$-valued gauge field $A^L_\mu+A^R_\mu$, seven scalar fields $Z^1,Z^2,Z^3,{\rm Re}Z^4$ and fermions which are now in the adjoint of the unbroken $U(N)$ gauge group. Furthermore since the vev eq.(\ref{vevmp}) is a maximally supersymmetric vacuum of the ABJM theory small fluctuations around it must have at least 12 supersymmetries. The only low energy theory that fits the description of these fluctuations is three-dimensional maximally supersymmetric Yang-Mills with gauge group $U(N)$ and this is indeed what one finds \cite{Mukhi:2008ux}. In particular the Yang-Mills coupling constant of this effective theory is $g_{YM} \propto v/k$ and hence this is weakly coupled as $k\to \infty$. Furthermore the higher derivative corrections are suppressed by powers of $v^{-2}$ and  are irrelevant as $v\to \infty$. 
 Thus we do indeed arrive at the correct weakly coupled D2-brane description. 

Finally we ask what the role of $l=|M-N|$ is? Note that if $l>0$ then parity is broken however the theory has the same moduli space as $l=0$ and hence probes the same spacetime metric. Thus the spacetime of these theories must break parity but still propagate in ${\mathbb C}^4/{\mathbb Z}_k$. It was argued in \cite{Aharony:2008gk} that $l$ corresponds to the amount of discrete torsion in the background supergravity four-form gauge field stength. In particular the ${\mathbb C}^4/{\mathbb Z}_k$ orbifold admits a degenerate  ${\mathbb Z}_k$ homology torsion 4-cycle\footnote{More accurately the AdS dual spacetime is $AdS_4\times S^7/{\mathbb Z}_k$ and   $H^4({\mathbb S}^7/{\mathbb Z}_k,{\mathbb Z})={\mathbb Z}_k$}. We can therefore turn on $l=0,1,2,..,k-1$ units of flux through the supergravity four-form. This will break parity but otherwise preserve supersymmetry and can be thought of as placing $l$ fractional M2-branes at the orbifold singularities. These are known as the ABJ models \cite{Aharony:2008gk}. Note that since $l\le k$ for this interpretation to make sense we require that the $U(N)\times U(N+l)$ theories do not exist for $l\ge k$, where $k$ is the level. Note that such theories are always strongly coupled since $M/k> l/k>1$ so there is no contradiction with the apparently good perturbative behaviour of the  weakly coupled theory deduced from the lagrangian.

\subsection{16 Supersymmetries Revisited}

For $k=1,2$ the $U(N)\times U(M)$ (with $M=N$ or $M=N+1$) vacuum moduli spaces are
\be
{\cal M}_{k=1} = \frac{({\mathbb R}^{8})^N}{S_N}\ ,\qquad {\cal M}_{k=2} = \frac{({\mathbb R}^{8}/{\mathbb Z}_2)^N}{S_N} \ ,
\ee
corresponding to $N$ M2-branes with transverse spaces ${\mathbb R}^{8}$ and ${\mathbb R}^{8}/{\mathbb Z}_2$ respectively. However in these cases the M2-branes should have maximal, 16 supersymmetries, certainly their moduli spaces do. But these are not manifest in the lagrangian. 

The solution to this paradox is that the theory is strongly coupled for $k=1,2$. Thus it is conceivable that the theory actually admits extra, hidden supersymmetries, at $k=1,2$. Indeed  \cite{Aharony:2008ugx} presents a chain of dualities that mapped their model to M2-branes moving in ${\mathbb R}^{8}/{\mathbb Z}_k$ which then implies that the extra supersymmetries must exist at $k=1,2$. It is beyond the scope of this review to give the details of how this works. We will simply say here that the extra supercurrents arise from so-called monopole or 't Hooft operators \cite{'tHooft:1977hy} which correspond to changing the topological properties of the fields in the path integral. These operators, although local,  cannot be expressed in terms of the fields that appear in the lagrangian. For more details of how this works see \cite{Klebanov:2008vq,Hosomichi:2008ip,Benna:2009xd,Gustavsson:2009pm,Kwon:2009ar,Kim:2009ia,Bashkirov:2010kz,Kapustin:2010xq,Martinec:2011}.

Now that we have discussed the spacetime interpretation of the ABJM and ABJ models, let us return to the BLG model that we first constructed. This is a special case of the general ${\cal N}=6$ theories where we take the 3-algebra generated by $2\times 2$ matrices. Choosing the basis 
\begin{equation}
T^a =
\left\{-\frac{i}{\sqrt{2}}\sigma_1,-\frac{i}{\sqrt{2}}\sigma_2,-\frac{i}{\sqrt{2}}\sigma_3,\frac{1}{\sqrt{2}}1_{2
\times 2}\right\}\ ,
\end{equation}
where $a=1,2,3,4$, $\sigma_i$ are the Hermitian Pauli matrices: $\sigma_i\sigma_j = \delta_{ij}
+i\epsilon_{ijk}\sigma^k$, we find that eq.(\ref{6triplep}) gives
\be
f^{abcd} = \frac{2\pi}{k}\varepsilon^{abcd}\ ,\qquad h^{ab}=\delta^{ab}\ ,
\ee 
where $h^{ab} = {\rm tr}(T^\dag_a,T^b)$. This is the totally anti-symmetric 3-algebra we saw in the BLG theory. Indeed one can check that the ${\cal N}=6$ lagrangian is just the ${\cal N}=8$ lagrangian written in complex form and hence has 16 supersymmetries. Thus the BLG lagrangian with structure constants $f^{abcd} = \frac{2\pi}{k}\varepsilon^{abcd}$ corresponds to the $su(2)\oplus su(2)$  ${\cal N}=6$ lagrangian.

Is there a role for the maximally supersymmetric $su(2)\oplus su(2)$ theories? They seem to be very similar to the $su(2)\oplus su(2)\oplus u(1)$ theories, especially once one realizes, following the previous discussion,  that the main role of the $u(1)$ factor is simply to impose the ${\mathbb Z}_k$ orbifold. To answer this we should look at their moduli space (for a more detailed study see \cite{Lambert:2010ji}). The calculation is similar to the ${\cal N}=6$ discussion but with a few important differences. In particular the moduli space effective action is
\be
{\cal L}_{vacuum} = -\sum^2_{i=1}D_\mu z^A_i D^\mu \bar z_{Ai} + \frac{k}{4\pi} \varepsilon^{\mu\nu\lambda} B_\mu\partial_\nu A_\lambda\ ,
\ee
where now $D_\mu z^A_1=\partial_\mu z^A_1 -i A_\mu z^A_1 $ and $D_\mu z^A_2=\partial_\mu z^A_2 -i A_\mu z^A_2 $. Note that even though we have $N=2$ there  is just one pair of $U(1)$ gauge fields, $A_\mu = A_\mu^{3L}-A_\mu^{3R}$, $B_\mu = A_\mu^{3L}+A_\mu^{3R}$, arising from the $\sigma_3$ components of the $su(2)\times su(2)$ gauge fields. Furthermore the two moduli $z^A_1$ and $z^A_2$ are oppositely charged. 

How is the previous argument altered? Firstly one still has the identification 
\be
z^A_1\leftrightarrow z^A_2\ ,
\ee
which arises from the adjoint action of the guage group and hence is oblivious to any $u(1)$ factors. But next we need to discuss the action of the
single $u(1)$ generator (recall there are two such $u(1)$'s in the $U(2)\times U(2)$ theory) which leads to the spacetime orbifold. Here we need to distinguish between the two global gauge groups: $SU(2)\times SU(2)$ and $(SU(2)\times SU(2))/{\mathbb Z}_2$. The reason is that the $F^{L/R}=dA^{L/R}$ field strengths obey  the standard Dirac quantization condition eq.(\ref{Dirac}) for $SU(2)\times SU(2)$  but this is modified to allow for half-integer fluxes in the case of $(SU(2)\times SU(2))/{\mathbb Z}_2$ (\eg\ see \cite{Lambert:2010ji}). This will lead to different moduli spaces. In particular one finds that the action of the orbifold becomes
\bea
z^A_1 \cong e^{\pi i/k}z^A_1 \qquad z^A_2 \cong e^{-\pi i/k}z^A_2 &&\qquad SU(2)\times SU(2) \\
z^A_1 \cong e^{2\pi i/k}z^A_1 \qquad z^A_2 \cong e^{-2\pi i/k}z^A_2&&\qquad (SU(2)\times SU(2))/{\mathbb Z}_2\ . \nonumber
\eea
In general this leads to the moduli spaces \cite{Lambert:2008et,VanRaamsdonk:2008ft,Lambert:2010ji}
\bea
{\cal M} &= & {\mathbb R}^{16}/D_{2k} \qquad SU(2)\times SU(2) \\
{\cal M} &= & {\mathbb R}^{16}/D_{k}\qquad (SU(2)\times SU(2))/{\mathbb Z}_2\ . \nonumber
\eea
where $D_{n} ={\mathbb Z}_2 \ltimes {\mathbb Z}_n$ is the dihedral group. There are three interesting cases to consider (here the subscripts give the level $k$ of the gauge group):
\begin{itemize}
\item $(SU(2)_1\times SU(2)_{-1})/{\mathbb Z}_2$:\qquad \ ${\cal M} = ({\mathbb R}^{8}\times {\mathbb R}^{8})/{\mathbb Z}_2$\ .
\item $ SU(2)_2\times SU(2)_{-2} $ : \qquad\qquad ${\cal M} = ({\mathbb R}^{8}/{\mathbb Z}_2\times {\mathbb R}^{8}/{\mathbb Z}_2)/{\mathbb Z}_2$\ .
\item $(SU(2)_4\times SU(2)_{-4})/{\mathbb Z}_2$  :\qquad ${\cal M} = ({\mathbb R}^{8}/{\mathbb Z}_2\times {\mathbb R}^{8}/{\mathbb Z}_2)/{\mathbb Z}_2$ \ .
\end{itemize}
These all correspond to two indistinguishable objects in ${\mathbb R}^{8} $ or ${\mathbb R}^{8}/{\mathbb Z}_2$ and therefore  nicely agree with the ABJM and ABJ theories.
Therefore it seems reasonable to conjecture that the following ABJM/ABJ theories are dual to BLG models  \cite{Lambert:2010ji,Bashkirov:2011pt}:
\begin{itemize}
\item $U(2)_1\times U(2)_{-1}$ is dual to $(SU(2)_1\times SU(2)_{-1})/{\mathbb Z}_2$.
\item $U(2)_2\times U(2)_{-2}$ is dual to $SU(2)_2\times SU(2)_{-2}$.
\item $U(2)_2\times U(3)_{-2}$ is dual to $(SU(2)_4\times SU(2)_{-4})/{\mathbb Z}_2$ .
\end{itemize}
These dualities have also been  tested more non-trivially by showing that their superconformal indices agree  \cite{Bashkirov:2011pt}  or in the case of the first duality by explicitly intergrating out the $u(1)$ gauge field of ABJM \cite{Lambert:2010ji}\footnote{More generally one finds that, if $k$ and $N$ are relatively prime, then the $U(N)\times U(N)$ ABJM theory is a ${\mathbb Z}_k$ orbifold of the $(SU(N)\times SU(N))/{\mathbb Z}_N$ $ {\cal N}=6$ theory.}.

\section{Future Directions and Summary}

\subsection{M5-branes}

There remains one last important class of conformal field theories that are predicted to exist by String Theory and M-Theory. Namely what about the low energy dynamics of multiple M5-branes? Such a theory should propagate in 6 spacetime dimensions, be conformal and admit maximal supersymmetry. In addition it should have five dynamical scalar fields $X^I$. Such a multiplet exists and is called the $(2,0)$ self-dual tensor multiplet and the worldvolume quantum field theory is often called simply   the $(2,0)$ theory. The multiplet is chiral because the M5-brane preserves supersymmetries that satisfy $\Gamma_{012345}\epsilon=\epsilon$. Its name derives from the fact that the remaining three bosonic degrees of freedom come from a self-dual three-form $H_{\mu\nu\lambda}$. The understanding of this theory is still an open question. Apart from its interest within M-theory it would be the first example of well-defined quantum field theory above four dimensions and has many rich features. It has also been related to the geometric Langlands programme in mathematics \cite{Witten:2009at}.

Following the flow of this review it is worth mentioning what happens if we apply the ideas that we have used above to this theory as was done in \cite{Lambert:2010wm}.  What do we need to look for? The theory should have a self-dual three-form $H_{\mu\nu\lambda}{}_a$, five scalars $X^I_a$ and fermions $\Psi_a$ that satisfy $\Gamma_{012345}\Psi_a=-\Psi_a$. To begin with we consider the free supersymmetry transformations:
\bea\label{Habelian}
\delta X^I_a &=& i \bar \epsilon \Gamma^I \Psi_a\cr
\delta  \Psi_a &=& \Gamma^\mu \Gamma_I \partial_\mu X^I_a \epsilon + \frac{1}{  3!}\frac{1}{2} \Gamma^{\mu\nu\lambda}H_{\mu\nu\lambda}{}_a\epsilon\; \cr
\delta H_{\mu\nu\lambda}{}_a &=& 3i \bar \epsilon \Gamma_{[\mu\nu} \partial_{\lambda]}\Psi_a \; .
\eea
To generalize this to these to an interacting theory we need to add non-linear terms to the right-hand-side of $\delta \Psi_a$. 
When doing so we are required to ensure that $\Gamma_{012345}\Psi_a=-\Psi_a$ and this means that there must be an odd number of $\Gamma_\mu$'s on the right-hand-side and hence an odd number of $\mu$ indices on the fields to soak up the free indices. There already are terms involving $D_\mu X^I_a$ and $H_{\mu\nu\lambda\; a}$ and so it would seem that we need to invent a new field with indices that can contract with the odd number of  $\Gamma_\mu$'s, since including higher powers of $D_\mu X^I_a$ and $H_{\mu\nu\lambda\; a}$ would lead to a higher-derivative theory. The simplest case is to include a new vector-like field $C^\mu_a$. Just as before turning on interactions will  lead  to a gauge symmetry and so we will also need to add a gauge field $A_\mu{}^b{}_a$. Putting these together and closing the algebra leads to the following set of supersymmetry transformations:
\bea\label{ansatz}
\delta X^I_a &=& i \bar \epsilon \Gamma^I \Psi_a\nn\\
\delta \Psi_a &=& \Gamma^\mu \Gamma^I D_\mu X_a^I \epsilon + \frac{1}{ 3!}\frac{1}{2} \Gamma_{\mu\nu\lambda}H_a^{\mu\nu\lambda}\epsilon- \frac{1}{2}\Gamma_\lambda \Gamma^{IJ} C^\lambda_b X^I_c X^J_d {f^{cdb}}_a\epsilon\nn\\
\delta H_{\mu\nu\lambda\; a} &=& 3 i \bar \epsilon \Gamma_{[\mu\nu}D_{\lambda]} \Psi_a +  i\bar \epsilon \Gamma^I \Gamma_{\mu\nu\lambda\kappa}C^\kappa_{b} X^I_c \Psi_d{f^{cdb}}_a \nn\\
\delta    A_{\mu\;a}^{\;b} &=& i \bar \epsilon \Gamma_{\mu\lambda} C^\lambda_c \Psi_d {f^{cdb}}_a\nonumber \\
\delta C^\mu_a  &=& 0\;,
\eea  
Which close onto the following equations of motion:
\bea\label{eq1}
0 &=&D^2 X_a^I -\frac{i}{2}\bar\Psi_cC^\nu_b\Gamma_\nu\Gamma^I \Psi_d f^{cdb}{}_a - C^\nu_b C_{\nu g} X^J_cX^J_eX^I_f f^{efg}{}_{d}f^{cdb}{}_a
 \cr
0 &=& D_{[\mu}H_{\nu\lambda\rho]\;a}+\frac{1}{4}\epsilon_{\mu\nu\lambda\rho\sigma\tau}C^\sigma_b X^I_cD^\tau X^I_df^{cdb}{}_a + \frac{i}{8}\epsilon_{\mu\nu\lambda\rho\sigma\tau}C^\sigma_b \bar\Psi_c\Gamma^\tau \Psi_d f^{cdb}{}_a \cr
0 &=& \Gamma^\mu D_\mu\Psi_a+X^I_cC^\nu_B\Gamma_\nu\Gamma^I\Psi_d f^{cdb}{}_a \cr
0&=&   F_{\mu\nu}{}^b{}_a + C^\lambda_cH_{\mu\nu\lambda\; d}f^{cdb}{}_a\cr
0 &=& D_\mu C^\nu_a = C^\mu_cC^\nu_df^{bcd}{}_a \cr
0 &=& C^\rho_cD_\rho X^I_d f^{cdb}{}_a = C^\rho_c D_\rho \Psi_d f^{cdb}{}_a =C^\rho_cD_\rho H_{\mu\nu\lambda\;a} f^{cdb}{}_a \;.
\eea
Note the appearance again of the 3-algebra structure constants. These are anti-symmetric in all the first three indices, just as in the M2-brane lagrangian above with 16 supersymmetries, and must satisfy the fundamental identity eq.(\ref{FIa}). However it need not have a positive definite metric.  Similar combinations of one-form gauge field and two-form potential (rather than the 3-form field strength used here) have appears in the mathematical literature of  so-called Lie 2-groups (for a review of these topics see \cite{Baez:2010ya} and \cite{Hofman:2002ey} for an early application to M5-branes). Indeed the above system can be very naturally recast in such a formalizism \cite{Palmer:2012ya}.

Although this does indeed serve as a representation of the $(2,0)$ superalgebra, at first sight  it seems to be something of a let-down since the final equation of motion tells us that the derivatives of the fields along the direction determined by $C^\mu_a$ (which is covariantly constant and hence non-dynamical) vanish. Thus we appear to have a five-dimensional system, dressed-up to look six-dimensional. Although we should be careful. If one looks a the superalgebra more carefully one finds that all components of the six-dimensional conserved quantities  are   non-vanishing \cite{Lambert:2011gb}, for example:
\bea
\nonumber
T_{\mu \nu} &=& D_\mu X^I_a D_\nu X^{Ia} - \frac{1}{2} \eta_{\mu \nu} D_\lambda X^I_a D^\lambda X^{Ia} \\
	&&+ \frac{1}{4} \eta_{\mu \nu} C^\lambda_b X^I_a X^J_c C_{\lambda g} X^I_f X^J_e f^{cdba} f^{efg}{}_d + \frac{1}{4} H_{\mu \lambda \rho \; a} H_{\nu}{}^{\lambda \rho \; a} \\
	&&- \frac{i}{2} \bar{\Psi}_a \Gamma_\mu D_\nu \Psi^a + \frac{i}{2} \eta_{\mu \nu} \bar{\Psi}_a \Gamma^\lambda D_\lambda \Psi^a - \frac{i}{2} \eta_{\mu \nu} \bar{\Psi}_a C^\lambda_b X^I_c \Gamma_\lambda \Gamma^I \Psi_d f^{abcd} \ .\nonumber 
\eea
In particular the momentum parallel to $C^\mu_a$ is given by the instanton number of the gauge fields  \cite{Rozali:1997cb}
\be
\frac{1}{8\pi^2} {\rm tr}\int   F\wedge F \in {\mathbb Z}\ ,
\ee 
which is discrete but certainly not vanishing. 

What exactly do we get from these equations? Well if we chose $C^\mu_a$ to be spacelike, say $C^\mu_a = g^2\delta^\mu_5\delta_a^\star$ where $\star$ is some direction in the 3-algebra then the equations reduce to those of five-dimensional maximally supersymmetric Yang-Mills with gauge algebra generated by $T^a$, $a\ne \star$ and Lie Bracket $[T^a,T^b] = ig^2f^{\star ab}{}_c T^c$. This is the theory on D4-branes, which corresponds to M5-branes wrapped on a circle. Comparing the instanton number with the Kaluza-Klein momentum one sees that the circle has radius $g^2/4\pi^2$. We could also examine the moduli space of vacua which is of course the same as that of $N$ D4-branes in String Theory - but then this is the same as what is expected of $N$ M5-branes in M-Theory.
Furthermore, there have been recent conjectures that the M5-brane theory on a circle is exactly five-dimensional maximally supersymmetric Yang-Mills \cite{Douglas:2010iu,Lambert:2010iw}.   Finally the role of instanton number as momentum seems to be an extension of the ABJM prescription that eleven-dimensional momentum is magnetic flux \cite{Lambert:2011eg}.

It is interesting to then ask what happens if we instead choose $C^\mu_a$ to be null: $C^\mu_a = g^2\delta^\mu_+\delta_a^\star$, where we have introduce lightcone coordinates $(x^+,x^-,x^i)$. Perhaps somewhat surprisingly, one can in fact show that the equations can all be reduced to one-dimensional motion on instanton moduli space \cite{Lambert:2011gb} with $x^-$ playing the role of `time'. The instanton number is now identified with the momentum along $x^+$. This system can be quantized and in fact leads to an old conjecture for the so-called discrete light-cone quantization of the  $(2,0)$ theory \cite{Aharony:1997th,Aharony:1997an}.

Thus we have lifted the D4-brane theory to look like a six-dimensional theory with $(2,0)$ supersymmetry, but compactified on a  circle, keeping the Kaluza-klein modes. In doing so we have unified two different descriptions of the M5-brane theory into a single system. How complete a theory of the M5-brane  eqs.(\ref{ansatz},\ref{eq1}) provides has yet to be determined but perhaps this system is not the disappointment that it first seemed to be. In addition it also seems to have nice applications to the worldvolume theories on other branes \cite{Honma:2011br,Kawamoto:2011ab}.

 Last, but not least, we note that some other proposals for defining the $(2,0)$ theory are given in \cite{ArkaniHamed:2001ie,Ho:2011ni,Chu:2011fd,Samtleben:2011fj,Chu:2012um}.

\subsection{Conclusion}

In this review we have attempted to describe how  String Theory and M-Theory predicts the existence of certain maximally supersymmetric interacting field theories, decoupled from gravity. Furthermore we have presented the explicit construction of these theories in the cases of all but the six-dimensional theory associated to M5-branes.  We have also concentrated on the most supersymmetric models of multiple M2-branes, namely those with 12 or 16 supersymmetries. However there has since been considerable work on constructing and understanding Chern-Simons-matter theories with various amounts of supersymmertries such as ${\cal N}=2,3,4,5$ ({\it e.g.} see  \cite{Gaiotto:2008sd,Cherkis:2008qr,Hosomichi:2008jd,Hosomichi:2008jb,Chen:2009cwa,Kim:2010kq,Bagger:2010zq,Palmkvist:2011aw}).

The M5-brane theory remains an important open problem. Constructing such a theory would also increase our knowledge of quantum field theory since, to date, there is no known description of a quantum field theory above four dimensions. The results so far suggest that five-dimensional maximally supersymmetric Yang-Mills could be a very interesting theory and possibly even well-defined \cite{Douglas:2010iu,Lambert:2010iw}. Understanding quantum field theory above four dimensions could also lead to more practical or phenomenological applications, in addition to presumably expanding our control over quantum field theory in general.

Perhaps we should return to our list in introduction. Needless to say the reader is surely no closer to achieving goal 6. We hope, however, that the reader is more convinced that the ideas which underlie M-Theory are fruitful and this is always a good sign. 

\section*{Acknowledgements}

I would like to thank J. Bagger, S. Mukhi, H. Nastase, C. Papageorgakis, P. Richmond, M. Schmidt-Sommerfeld and  D. Tong for their insightful and enjoyable collaboration.

\section*{Appendix: Conventions}

Let us summarize our conventions here. We use a mostly plus spacetime metic:
\be
\eta_{mn} = \left( \begin{array}{llll}-1 &&&\\ &1&&\\  & & \ddots &\\ &&&1\end{array} \right)\ ,
\ee 
where $m,n= 0,1..,10$ ($m,n=0,1,...,9$ in the case of String Theory). We will typically denote the worldvolume coordinates of a brane by 
$x^\mu$, $\mu = 0,1,2,...,p$ and the coordinates transverse the the brane by $x^I$, $I=p+1,..,10$ (or $I=p+1,...,9$ in the case of String Theory). The scalar fields that represent fluctuations of the brane in the transverse space are denoted by $X^I(x^\mu)$. 

We use a real (Majorana) basis of eleven-dimensional,  $32\times 32$, $\Gamma$-matrices such that
\be
\Gamma_m^\dag = \Gamma_m^T = -C\Gamma_m C^{-1}\ .
\ee
where $C = \Gamma_0$ and $m=0,...,10$. When we talk about ten-dimensional String Theory we can use the same $\Gamma$-matrices only we re-label $\Gamma_{10} = \Gamma_{11}$. 

We also assume that all spinors are real and anti-commuting: $\psi_1\psi_2=\psi_2\psi_1$.  We define
\be
\bar \epsilon = \epsilon^TC\ .
\ee 
Finally we assume that complex conjugation acts as $(\psi_1\psi_2)^*=\psi^*_2\psi^*_1$, for any two objects $\psi_1$ and $\psi_2$. This means that $\bar \epsilon \Gamma^m\Psi$ is imaginary.

\bibliographystyle{arnuke_revised}
\bibliography{OtherReview_v3}

\end{document}